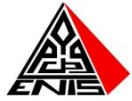

# MEMOIRE

*Présenté à*

**L'École Nationale d'Ingénieurs de Sfax**

*en vue de l'obtention du*

# MASTERE

*Dans la discipline Génie informatique*
*Nouvelles Technologies des Systèmes Informatiques Dédiés*

*Par*

**Imen SAYAR**
**(Maîtrise Informatique)**

# D'Event-B vers UML/OCL en passant par UML/EM-OCL

*Soutenu le 13 Février 2012, devant le jury composé de :*

| | | |
|---|---|---|
| M. | **Mohamed JMAIEL** (Professeur.ENIS) | *Président* |
| M. | **Kais HADDAR** (MA Habilité.FSS*)* | *Membre* |
| M. | **Mohamed Tahar BHIRI** (MA.FSS) | *Encadreur* |

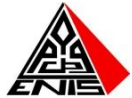

# D'Event-B vers UML/OCL en passant par UML/EM-OCL

## Imen SAYAR

**Résumé** : Pour surmonter les limites inhérentes aux deux approches classique et formelle de développement des logiciels complexes, nous avons proposé une approche hybride combinant l'approche formelle (Event-B) et l'approche classique (UML/OCL). Les phases en amont de notre approche comportent : Réécriture du cahier des charges, Stratégie de raffinement, Spécification abstraite et Raffinement horizontal. Nous avons montré la faisabilité de notre approche sur une étude de cas : Système de Clés Electroniques d'Hôtels (SCEH). Le problème de passage du formel (Event-B) vers le semi-formel (UML/OCL) est traité en passant par notre extension à OCL (EM-OCL).

**Mots clés**: Approche classique, Approche formelle, Approche hybride, Event-B, UML/OCL, UML/EM-OCL

**Abstract:** To overcome the limitations of both approaches classical and formal for the development of complex software, we proposed a hybrid approach combining the formal approach (Event-B) and the classical approach (UML/OCL). Upstream phases of our approach include: Rewriting the requirements document, Refinement strategy, Abstract specification and Horizontal refinement. We have shown the feasibility of our approach on a case study: An Electronic Hotel Key System (SCEH). The problem of transition from the formal (Event-B) to the semi-formal (UML/OCL) is processed through our extension to OCL (EM-OCL).

# Sommaire

# Table des figures

# *Liste des tableaux*

# *Introduction*

Les logiciels et systèmes informatiques deviennent de plus en plus complexes. Ceci les rend très sensibles aux erreurs produites durant le cycle de vie d'un système. En effet, ces erreurs logicielles peuvent être introduites durant les phases initiales (cahier des charges, spécifications générales, spécifications détaillées), phases médianes (conception générale, conception détaillée) et phases finales (codage, intégration, maintenance). Une erreur produite lors des phases initiales a un prix plus élevé que celle produite lors des phases ultérieures [14].

Le génie logiciel offre des méthodes, techniques et outils permettant d'éviter les erreurs inhérentes au développement des logiciels et systèmes complexes. En effet, les méthodes formelles, les techniques de test, le model-checking et l'interprétation abstraite favorisent le développement des logiciels et systèmes **corrects** [24] [39].

Les processus de développement formels organisés autour des méthodes formelles comme B [4] et Event-B [3] apportent des réponses pertinentes à la problématique de construction des logiciels et systèmes complexes. De tels processus permettent de combattre les erreurs très tôt dès les phases initiales. En outre, grâce à la technique de raffinement [44], B et Event-B permettent l'élaboration incrémentale d'une spécification abstraite cohérente (modèle abstrait initial) et concrétiser pas à pas cette spécification abstraite jusqu'à la génération de code. Ceci autorise la fabrication des logiciels et systèmes **corrects par construction**. Plusieurs outils supportent les méthodes formelles tels que : générateur des obligations de preuves, prouveur automatique et interactif, animateur et model-checker [11]. Les deux premiers outils sont utilisés pour **vérifier** la cohérence et la correction de raffinement des modèles développés. Tandis que les deux autres outils sont utilisés pour **valider** [6] les modèles développés par les méthodes formelles.

Mais l'utilisation des méthodes formelles comme B et Event-B de bout en bout dans le cycle de vie d'un logiciel ou système se heurte, à notre avis, aux problèmes suivants :

(1) L'utilisateur final est pratiquement exclu. En effet, il ne peut pas participer à la validation des modèles formels développés exigeant des notations mathématiques rigoureuses. Afin de résoudre ce problème, des travaux permettant la traduction des modèles B vers des modèles UML existent [17].

(2) Le processus de raffinement de l'abstrait (modèle abstrait) vers le concret (modèle concret) n'est pas évident. Et il n'existe pas des guides méthodologiques standards à l'instar des patterns architecturaux, d'analyse et de conception [12] dans le monde OO (Orienté Objet).

(3) La maintenance évolutive ayant pour objectif l'intégration de nouveaux besoins exige souvent la révision des modèles formels déjà développés, la stratégie de raffinement adoptée et des efforts plus ou moins importants liés à l'activité de preuve.

Pour faire face à ces problèmes, nous proposons un processus de développement combinant le formel (Event-B) et le semi-formel (UML/OCL) en passant par notre extension UML/EM-OCL [40] [8]. Le processus de développement préconisé utilise Event-B que dans les phases initiales (cahier des charges, spécification générale et spécification détaillée) puis passe la main à l'approche par objets autour d'UML/OCL pour les autres phases médianes et finales. Le niveau intermédiaire UML/EM-OCL facilite la traduction systématique d'Event-B vers UML/OCL. Le processus de développement préconisé est illustré sur une étude de cas Systèmes de Clés Electroniques pour Hôtels décrite en [3].

Ce mémoire comporte quatre chapitres :

- Le premier chapitre donne une vue d'ensemble sur les processus de développement existants des logiciels et systèmes. En outre, il décrit et positionne le processus de développement préconisé combinant le formel et semi-formel.

- Le second chapitre présente d'une façon rigoureuse les constructions offertes par Event-B (machine abstraite, machine raffinée, contexte, langage logico-ensembliste, langage de substitutions, obligations de preuves et preuve interactive) permettant de développer des modèles corrects par construction en se servant de la technique de raffinement. En outre, un aperçu sur la plateforme RODIN [33] [34] supportant la méthode Event-B est fourni dans ce chapitre. Enfin, ce chapitre décrit l'extension mathématique d'OCL appelée EM-OCL issue de notre équipe.

- Le troisième chapitre propose une modélisation en Event-B de l'application Système de Clés Electroniques pour Hôtels (SCEH) en appliquant les phases initiales de notre processus de développement proposé dans le chapitre 1.

- Le dernier chapitre est consacré au passage du modèle Event-B de l'application SCEH vers une modélisation en UML/OCL en passant par un modèle pivot développé en UML/EM-OCL. Le modèle en Event-B à traduire en UML/OCL en passant par UML/EM-OCL décrit avec précision les propriétés et le fonctionnement de l'application SCEH. Il est obtenu en appliquant un raffinement horizontal [3] garantissant que toutes les propriétés issues du cahier des charges réécrites selon les recommandations préconisées dans [2] et [36] de l'application SCEH ont été prises en compte. En outre, ce dernier chapitre propose des règles de traduction systématique selon deux niveaux de modélisation : de la spécification Event-B vers un modèle UML/EM-OCL, puis du modèle UML/EM-OCL vers un modèle UML/OCL.

# Chapitre 1 : Vers un processus de développement hybride des logiciels

## 1.1 Introduction

Dans ce chapitre, nous abordons les processus de développement de logiciels. Dans un premier temps, nous présentons et analysons l'approche classique de développement de logiciels. Dans un deuxième temps, nous présentons et analysons l'approche formelle de développement de logiciels. Dans un troisième temps, nous proposons une approche hybride combinant les deux approches classique et formelle. L'approche formelle est utilisée pour les phases initiales (spécification) et l'approche classique est utilisée pour les phases médianes et finales (conception, codage et intégration). Enfin, nous appliquons l'approche hybride proposée à Event-B et UML/OCL.

## 1.2 Activité du développement de logiciels

Le développement de logiciels nécessite plusieurs types d'activités [19] :

- Définir ce qui sera développé
- Définir comment il sera développé
- Développer un des composants
- Assembler les composants
- Valider le logiciel

Ces activités produisent plusieurs types de documents (ou artefacts) tels que : cahier des charges, modèles, implémentation (ou code), programme (ou fichier exécutable), procédures de test et manuels d'utilisation.

L'organisation de ces activités et leur enchaînement définit le cycle de développement (ou cycle de vie) du logiciel. En génie logiciel, plusieurs cycles de vie sont utilisés afin de développer des logiciels plus ou moins de qualité tels que : cycle de vie en cascade [27], cycle de vie en V [28], cycle de vie en « spirale » [28], RUP (Rational Unified Process) [35] et MDA (Modelling Driven Architecture) [23].

## 1.3   Approche classique

L'approche classique pour le développement de logiciels comporte plusieurs phases : spécification, conception, codage et intégration.

L'enchaînement de ces phases est illustré par la **figure 1.1**. Plusieurs formalismes peuvent être utilisés par les différentes phases de l'approche classique. Par exemple, dans un développement Orienté Objet (OO) centré autour d'UML, on peut utiliser respectivement les diagrammes de cas d'utilisation et de séquences (pour la phase de spécification), et les diagrammes de classes, d'état-transition et d'activités (pour la phase de conception). En outre, dans un cadre OO, on peut réutiliser avec profit les patterns architecturaux [30], d'analyse et de conception [12] (pour la phase de conception) et les bibliothèques de classes [26] (pour les deux phases de codage et d'intégration). Mais, l'approche classique pour le développement de logiciels présente plusieurs insuffisances :

- Les cahiers des charges utilisés par l'approche classique sont très souvent difficiles à exploiter. Ils présentent des mécanismes de réalisation au détriment de **l'explicitation des propriétés** du futur logiciel. D'ailleurs, l'approche classique n'apporte pas des guides méthodologiques favorisant l'écriture des cahiers des charges de qualité [2].

- Les risques d'erreurs sont à toutes les phases de l'approche classique. En outre, ces erreurs sont souvent uniquement détectées dans les **phases finales** après la phase de codage et d'intégration.

- La détection des erreurs nécessite des tests très poussés : test unitaire, d'intégration et système [45]. Ces tests sont très coûteux et ne garantissent pas **la correction** du logiciel développé.

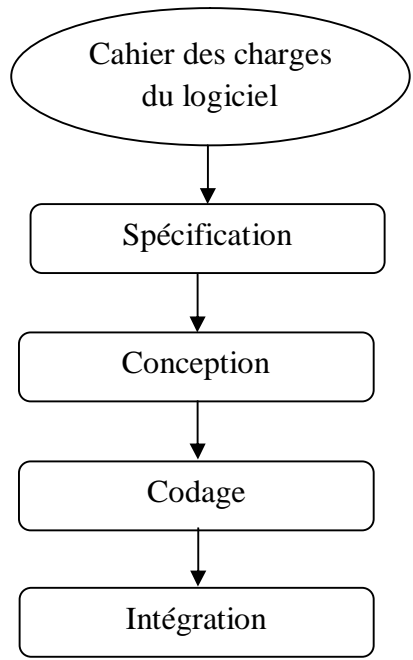

**Figure 1.1 :** *Approche classique de développement de logiciels*

## 1.4  Approche formelle

Une approche formelle pour le développement de logiciels est basée sur des méthodes formelles. Les deux éléments fondamentaux d'une méthode formelle sont : langage formel et processus formel de développement. Un langage formel est une notation mathématique **bien définie** aussi bien sur le plan syntaxique que sémantique. Un processus formel de développement permet le passage pas-à-pas de l'abstrait (quoi ou encore la spécification) vers le concret (comment ou encore l'implémentation) en garantissant que l'implémentation est **correcte par construction** vis-à-vis de sa spécification moyennant une activité de **preuve**.

A titre d'exemple, B [4] est une méthode formelle avec preuves englobant un langage logico-ensembliste et de substitutions généralisées et un processus formel de développement basé sur la technique de raffinement. Egalement, Event-B est une méthode formelle avec preuves (voir chapitre 2).

L'approche formelle pour le développement de logiciels comporte les phases suivantes : réécriture du cahier des charges, spécification abstraite, raffinement horizontal, raffinement vertical et génération de code. L'enchaînement de ces phases est illustré par la **figure 1.2**.

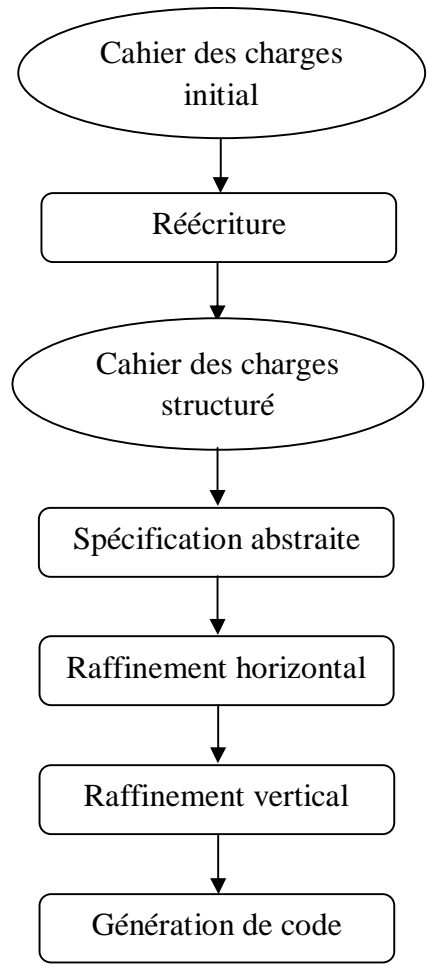

**Figure 1.2 :** *Approche formelle pour le développement de logiciels*

La phase de *réécriture du cahier des charges* a pour objectif de réécrire le cahier des charges initial de façon à mettre en exergue les propriétés du logiciel à développer. En effet, les cahiers des charges tels qu'ils sont élaborés aujourd'hui ne se prêtent pas bien à l'utilisation des méthodes formelles [2]. La phase de *spécification abstraite* produit un modèle initial très abstrait. Celui-ci est affiné pas-à-pas en tenant compte des propriétés explicitées par le cahier des charges de qualité. Le dernier modèle issu de la phase de *raffinement horizontal* intègre toutes les propriétés et tous les comportements du logiciel à développer. Ceci constitue la spécification du logiciel. La phase de *raffinement vertical* a pour objectif d'apporter une implémentation pas-à-pas à la spécification fournie par le modèle issu de la phase de *raffinement horizontal*. Ainsi, une succession de modèles de moins en moins abstraits est produite lors de cette phase. L'ultime modèle issu de la phase de *raffinement vertical* est transformé en code par la phase de *génération de code*.

L'approche formelle de développement de logiciels est supportée par une activité de preuve mathématique outillée (générateur d'obligations de preuves, prouveur, animateur et model-checker) afin de vérifier la cohérence des modèles produits et la correction de chaque étape de raffinement aussi bien pour le raffinement horizontal que vertical.

L'approche classique peine à développer des logiciels corrects. Tandis que, l'approche formelle permet l'obtention des logiciels corrects par construction. En effet, chaque modèle produit par cette approche est soumis à des preuves. Si une preuve échoue, alors le modèle est revu et corrigé. Puis, on crée un modèle plus affiné, auquel on applique des preuves. Et ainsi de suites jusqu'au modèle ultime transformé automatiquement en code grâce au générateur de code. En outre, contrairement à l'approche classique, la phase des tests est très simplifiée dans l'approche formelle. Enfin, les deux approches ont globalement un coût équivalent [1]. Dans l'approche formelle, le temps passé à élaborer les modèles successifs est rattrapé lors de la phase des tests qui prend très peu de temps, contrairement à l'approche classique [31].

Mais l'utilisation des méthodes formelles comme B et Event-B de bout en bout dans le cycle de vie d'un logiciel ou système se heurte, à notre avis, aux problèmes suivants[1] :

(1) L'acteur qui ne domine pas le processus formel utilisé est pratiquement exclu. En effet, il ne peut pas participer à la validation des modèles formels développés exigeant des notations mathématiques rigoureuses. Afin de résoudre ce problème, des travaux permettant la traduction des modèles B vers des modèles UML existent [17].

(2) Le processus de raffinement de l'abstrait (modèle abstrait) vers le concret (modèle concret) n'est pas évident. Et il n'existe pas des guides méthodologiques standards à l'instar des patterns architecturaux, d'analyse et de conception [12] dans le monde OO (Orienté Objet).

(3) La maintenance évolutive ayant pour objectif l'intégration de nouveaux besoins exige souvent la révision des modèles formels déjà développés, la stratégie de raffinement adoptée et des efforts plus ou moins importants liés à l'activité de preuve.

---

[1] Ces problèmes sont déjà signalés dans l'introduction de ce mémoire.

## 1.5 Approche hybride

Afin de développer des logiciels de qualité, nous préconisons une approche hybride combinant les deux approches précédemment présentées à savoir l'approche classique (voir **1.3**) et l'approche formelle (voir **1.4**). Les différentes phases de l'approche hybride proposée sont illustrées par la **figure 1.3**. Les phases initiales de l'approche hybride sont confiées à l'approche formelle afin de dériver un modèle formel explicitant les propriétés et comportements du futur logiciel. Pour y parvenir, on fait appel aux trois phases *Réécriture*, *Spécification abstraite* et *Raffinement horizontal* venant de l'approche formelle. Ensuite, on applique les trois phases *Conception*, *Codage* et *Intégration* venant de l'approche classique.

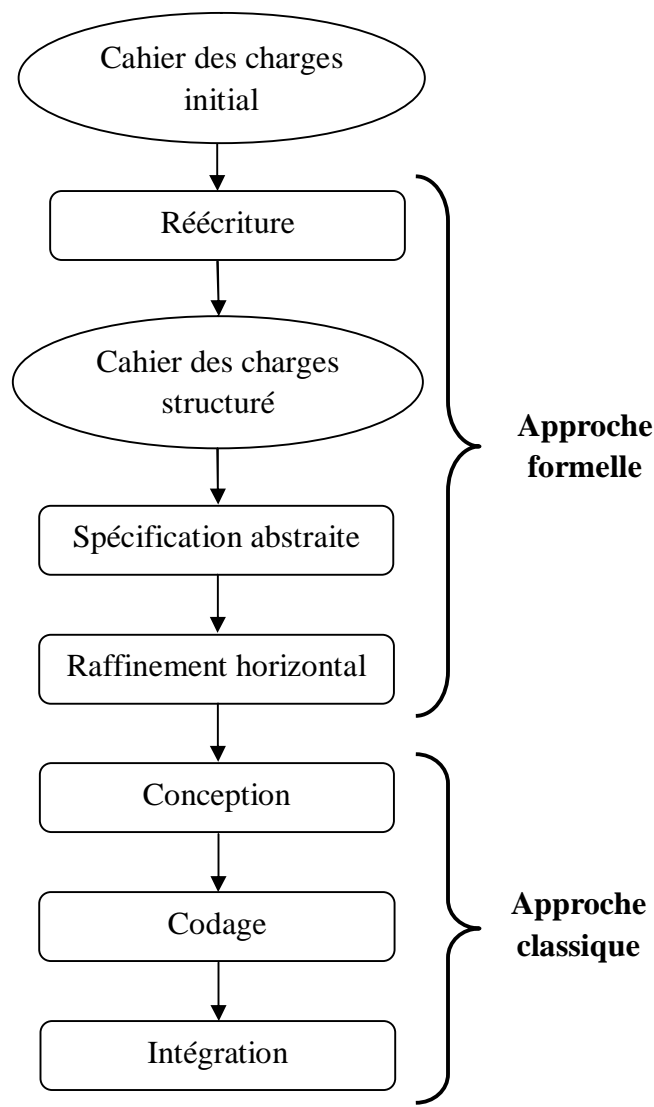

**Figure 1.3 :** *Approche hybride de développement de logiciels*

L'approche hybride permet de faire face aux défauts de deux approches classique et formelle signalées et expliquées dans les deux sections **1.3** et **1.4**. En effet, cette approche possède les mérites suivants :

- Elle garantit l'élaboration d'un modèle formel **cohérent** traduisant l'énoncé informel du cahier des charges initial. Ainsi, les risques d'erreurs liées à la phase *Spécification* de l'approche classique sont bel et bien écartés.

- Elle autorise la participation des acteurs non forcément experts dans les méthodes formelles à partir de la phase de *Conception*.

- Elle favorise la réutilisation des solutions conceptuelles et d'implémentation existantes, par exemple dans un développement OO : patterns de conception et bibliothèques de classes. Ceci facilite la maintenance évolutive du logiciel développé.

- Le modèle formel traduisant la spécification du futur logiciel peut être utilisé avec profit afin de faciliter la phase de tests.

Mais, l'approche hybride pose un nouveau problème : le passage du formel vers les formalismes utilisés par l'approche classique.

Dans la section suivante, nous apportons une solution à ce problème de passerelle dans le cadre d'Event-B (formel, approche formelle) et UML (semi-formel, approche classique).

## 1.6 Approche hybride appliquée à Event-B et UML

### 1.6.1 Présentation

Afin d'expérimenter l'approche hybride de développement de logiciels présentée dans la section **1.5**, nous avons retenu Event-B et UML. La méthode formelle Event-B est utilisée en amont pour les deux phases *Spécification abstraite* et *Raffinement horizontal*. Tandis que UML est utilisé notamment pour la phase de *Conception* en se servant du paradigme orienté objet. Dans notre approche hybride appliquée à Event-B et UML, les modèles traités sont, soit des modèles formels en Event-B soit des modèles semi-formels en UML. Pour apporter plus de précisions aux modèles UML, nous faisons appel au langage OCL. Le problème de passerelle entre Event-B et UML/OCL est traité en se servant de notre extension à OCL appelée **EM-OCL** [8] [40]. La **figure 1.4** donne les différentes phases de l'approche hybride appliquée à Event-B et UML/OCL. Dans la suite, nous allons détailler ces différentes phases.

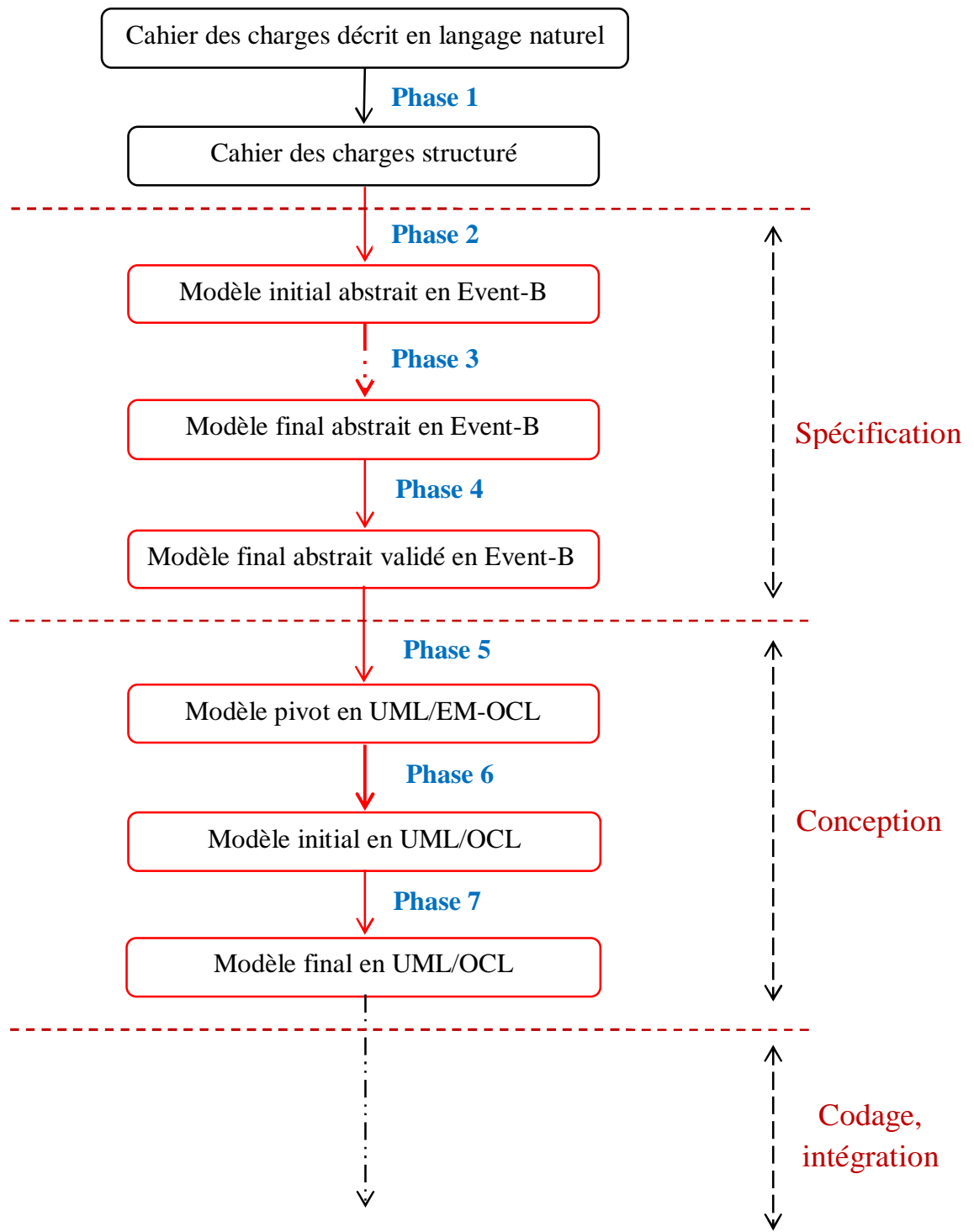

**Figure 1.4 :** *Approche hybride appliquée à Event-B et UML/OCL*

### 1.6.2 Explication

Dans cette section, nous décrivons l'apport de chaque phase formant un cycle de développement basé sur une approche hybride appliquée à Event-B et UML/OCL.

- ***Phase 1 : Restructuration du cahier des charges***

    Les cahiers des charges tels qu'ils sont élaborés aujourd'hui ne se prêtent pas bien à l'utilisation des méthodes formelles. En effet, ils ne peuvent pas être exploités judicieusement dans une optique de développement incrémental des logiciels en utilisant la technique de raffinement. Autrement dit, le spécifieur ne peut pas élaborer sa stratégie de raffinement en se référant explicitement au cahier des charges.

    Inspiré de l'organisation des documents en mathématiques, l'idée de cette phase consiste à réorganiser le cahier des charges en deux textes séparés décrits comme suit [3]:

    - Un <u>texte référentiel</u> : ce type de texte rassemble les contraintes et les propriétés essentielles du système à modéliser. Il doit être formé de courtes phrases écrites en langage naturel, simples à comprendre et libellées avec des étiquettes numérotées pour garantir la traçabilité. Une étiquette peut appartenir aux ensembles suivants :

        - **FUN** : pour les contraintes fonctionnelles,
        - **ENV** : pour les contraintes d'environnement c'est-à-dire liées aux équipements,
        - **SAF** : pour les propriétés favorisant la sécurité du système,
        - **DEL** : pour les contraintes liées aux délais,

    - Un <u>texte explicatif</u> : il englobe une description de tous les détails du système et sert à comprendre les exigences de l'application et ses détails. Une fois maîtrisé, ce genre de texte devient moins important que le texte référentiel qui est utilisé comme référence de l'application.

    La **figure 1.5** décrit une analogie entre la présentation d'un théorème dans un document mathématique et la structure d'un cahier des charges structuré.

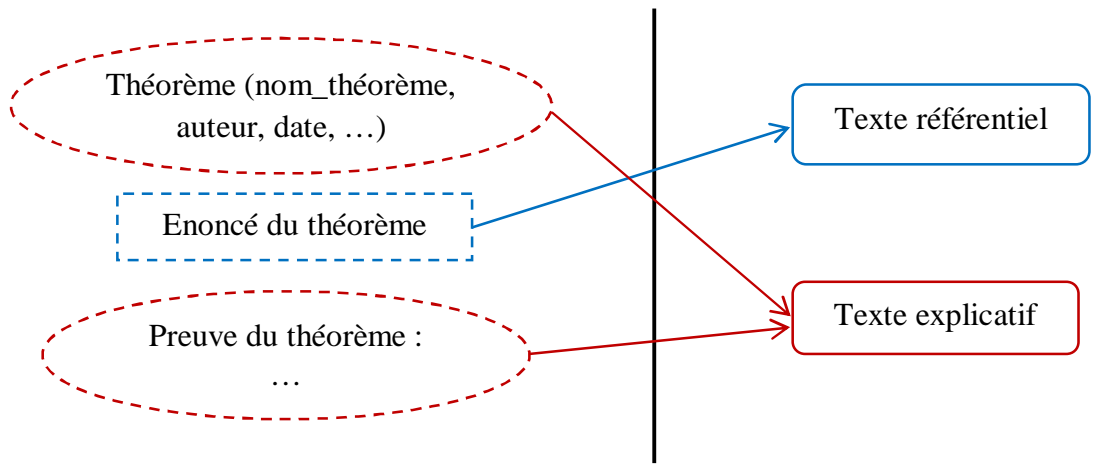

**Figure 1.5 :** *Analogie entre description mathématique d'un théorème et organisation d'un cahier de charges*

Le document issu de la phase 1 appelé **cahier des charges structuré** joue un rôle capital dans le processus de développement proposé. En effet, un tel document est déterminant pour l'élaboration d'une stratégie de raffinement horizontal permettant à terme d'établir une spécification **cohérente** et **complète** englobant toutes les propriétés et les comportements du futur logiciel.

- *Phase 2 : Formalisation du cahier des charges*

    Cette phase a pour objectif d'établir un modèle initial abstrait en Event-B décrivant les fonctions principales du futur système. Un tel modèle est souvent basé sur un petit nombre de propriétés et contraintes issues du cahier des charges structuré produit par la **phase 1**. Le modèle initial abstrait en Event-B produit par cette phase doit être soumis à des preuves au sens de la théorie Event-B : préservation des propriétés de sûreté et de vivacité (voir chapitre 2). En cas d'échec –des preuves non déchargées-, il doit être revu et corrigé jusqu'à l'obtention d'un modèle prouvé.

    Avant d'entamer cette phase, il est intéressant d'établir explicitement la *stratégie de raffinement horizontal* à suivre. Celle-ci définit l'ordre dans lequel on va extraire et formaliser les propriétés et contraintes du futur système. Le choix d'une stratégie de raffinement conditionne la qualité des modèles obtenus et par conséquent la réussite du processus de développement formel. Ceci constitue un des goulots d'étranglements de l'approche formelle. Des travaux permettent de proposer des stratégies de raffinement

plus ou moins « optimales » liées à des classes d'application commencent à apparaître [36].

- *Phase 3 : Raffinement du modèle initial abstrait*

　　La phase 3 produit à terme un modèle abstrait final en Event-B par des raffinements successifs en intégrant pas-à-pas des propriétés et des contraintes bien référencées issues du cahier des charges structuré selon la stratégie de raffinement adoptée. Chaque modèle affiné est soumis à des preuves mathématiques liées à sa cohérence et sa correction vis-à-vis de son modèle abstrait. Ceci est illustré par la **figure 1.6**.

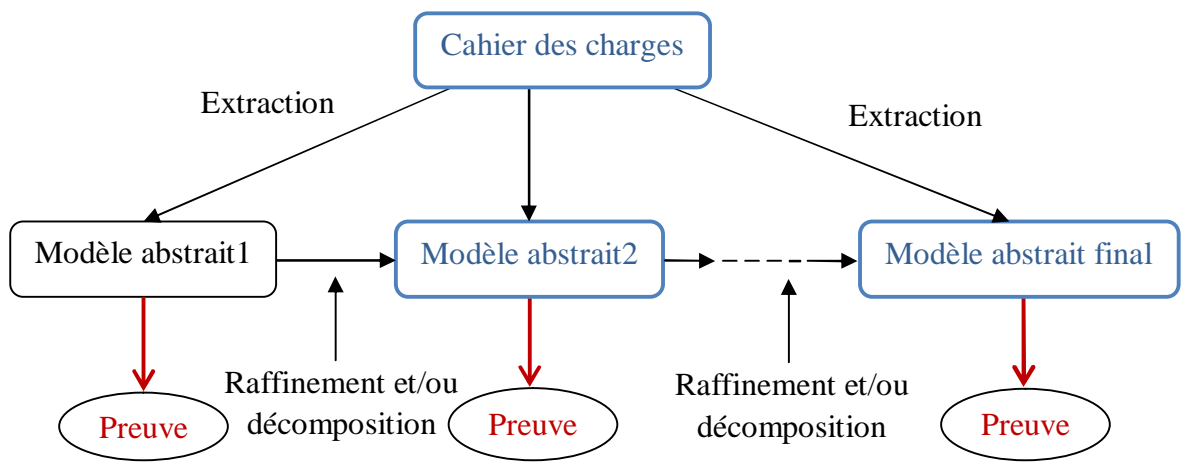

**Figure 1.6 :** *La macro-architecture de la phase 3*

- *Phase 4 : Validation des modèles Event-B :*

　　Un modèle Event-B prouvé n'est pas forcément correct. Bien que le cahier des charges structuré réduise d'une façon considérable les risques d'erreurs, il ne peut pas écarter définitivement les oublis des besoins. Pour faire face à ces oublis, une activité de validation des modèles Event-B s'impose en se servant des outils d'animation associés à Event-B tels que Brama [9] et ProB [21]. En outre, l'animation des modèles Event-B permet à l'utilisateur de participer très tôt à la validation [46] de la spécification de son futur logiciel. De plus, le model-checker peut être utilisé pour assister le spécifieur à décharger des grosses formules liées aux propriétés de vivacité (voir chapitre 2).

- *Phase 5 : Construction d'un modèle UML/EM-OCL à partir d'une spécification Event-B prouvée et validée*

    Le modèle Event-B –décrivant une spécification cohérente et validée- issu de la **phase 4** est traduit plus ou moins systématiquement en UML/EM-OCL [8] [40]. La traduction d'Event-B vers UML/EM-OCL est relativement simple car ils partagent en commun des concepts mathématiques tels que : paire, relation et fonction. Ainsi, le langage UML/EM-OCL joue le rôle du langage pivot entre Event-B et UML/OCL.

- *Phase 6 : Du modèle UML/EM-OCL vers un modèle UML/OCL*

    Le modèle en UML/EM-OCL issu de la phase précédente est transformé en UML/OCL en utilisant des règles de transformation établies par nous-mêmes (voir chapitre 4). Le modèle UML/OCL produit par cette phase traduit les concepts métier de l'application.

- *Phase 7 : Intégration des décisions de conception et d'implémentation :*

    Le modèle UML/OCL issu de la **phase 6** traduit les concepts métier du futur logiciel. La phase 7 a pour objectif d'intégrer respectivement des choix ou des décisions de conception et d'implémentation en se servant des patterns de conception de GoF [12] et des bibliothèques de classes.

    Le modèle UML/OCL issu de la **phase 7** est le modèle ultime et par conséquent il peut être concrétisé en appliquant les deux phases de *Codage* et d'*Intégration*. Enfin, la phase des tests (unitaires, d'assemblage et système) valide le logiciel obtenu. Notons au passage qu'il est possible d'utiliser avec profit le modèle final abstrait validé en Event-B issu de la **phase 4** afin de générer des données de tests [5].

## 1.7  Conclusion

Après avoir évalué les deux approches classique et formelle, nous avons proposé une approche hybride combinant ces deux approches. Ceci permet de faire face aux défauts inhérents à ces deux approches. Nous avons montré l'applicabilité de l'approche hybride proposée sur les deux formalismes Event-B (approche formelle) et UML/OCL (approche classique, cadre OO). Nous avons proposé une solution basée sur notre extension UML/EM-OCL pour le problème de passerelle entre Event-B et UML/OCL.

Dans le chapitre suivant, nous allons présenter les aspects fondamentaux de la méthode formelle Event-B et UML/EM-OCL utilisé comme pivot entre Event-B et UML/OCL.

# Chapitre 2 : Event-B et EM-OCL

## 2.1 Introduction

Ce chapitre comporte deux sections. La première section établit une synthèse sur la méthode formelle Event-B. Tous les aspects fondamentaux d'Event-B sont décrits d'une façon assez détaillée : contexte, machine abstraite, machine affinée, événement, obligations de preuves, types de raffinement et propriété de vivacité. La deuxième section établit également une synthèse sur le langage UML/EM-OCL issu de notre équipe. Celui-ci sera utilisé (voir chapitre 4) comme langage pivot entre Event-B et UML/OCL.

## 2.2 La méthode formelle Event-B

La méthode Event-B [18] [3] est une extension de B [4] qui permet la spécification des systèmes réactifs, des algorithmes séquentiels, concurrents et distribués. Cette méthode utilise des approches mathématiques [38] basées sur la théorie des ensembles et la logique de prédicats [10].

### 2.2.1 Logique et théorie des ensembles

Le langage logico-ensembliste d'Event-B est basé sur la logique classique du premier ordre et la théorie des ensembles.
Les symboles utilisés pour exprimer des prédicats logiques sont :

$\top$ vrai

$\bot$ faux

$P \wedge Q$ conjonction

$P \vee Q$ disjonction

$\neg P$ négation

$P \Rightarrow Q$ implication

P⇔Q équivalence

∀ $x_1, x_2, ..., x_n$ . $P(x_1, x_2, ..., x_n)$ quantification universelle

∃ $x_1, x_2, ..., x_n$ . $P(x_1, x_2, ..., x_n)$ quantification existentielle

Le langage logico-ensembliste d'Event-B supporte les notations ensemblistes usuelles comme :

- l'inclusion (⊆)
- l'inclusion stricte (⊂)
- l'union (∪)
- l'intersection (∩)
- la différence d'ensemble (\)
- l'ensemble vide (∅ ou {})
- l'ensemble des parties non vides (ℙ1)
- l'ensemble des parties (ℙ)
- les ensembles énumérés {$x_1, x_2, ..., x_n$}

En outre, le langage logico-ensembliste d'Event-B supporte les concepts mathématiques couple, relation et fonction. Les concepts usuels sur les relations (s->t) sont définis en Event-B tels que :

- domaine (dom)
- codomaine (ran)
- relation d'identité (id)
- relation réciproque ($r^{-1}$)
- image d'un ensemble par une relation (r[s])
- et la composition (r;q)

De plus, le langage logico-ensembliste d'Event-B propose des opérateurs pour restreindre les relations :

- la restriction sur le domaine (S◁r)
- l'anti-restriction pour enlever des éléments du domaine (S⩤r)
- la corestriction qui est une restriction sur le codomaine (r▷T)
- l'anti-corestriction pour enlever des éléments du codomaine (r⩥T)
- et la surcharge permet d'obtenir une relation r⩥p à partir de deux autres relations r et p.

Les fonctions sont considérées comme des relations particulières. On distingue :
- injective partielle (s⤔t)
- surjective partielle (s⤀t)
- totale (s→t)
- totale injective (s↣t)
- totale surjective (s↠t)
- et totale bijective (s⤖t)

Enfin, les couples en Event-B sont notés sous la forme x↦y.

Les constructions offertes par le langage logico-ensembliste permettent de typer les ensembles, les constantes et les variables décrivant la partie statique d'un modèle Event-B. En outre, elles sont utilisées pour décrire les propriétés invariantes d'un modèle Event-B sous forme des prédicats logiques. Enfin, elles décrivent les axiomes et théorèmes des modèles Event-B.

### 2.2.2 Modèles Event-B

Le modèle est le premier concept d'Event-B. Il est composé d'un ensemble de machines et de contextes. Une machine Event-B contient deux parties : statique et dynamique. La partie statique comporte les variables modélisant l'état du système. Ces variables sont typées et peuvent avoir des propriétés invariantes décrites par des prédicats logiques. La partie dynamique comporte des évènements permettant d'agir sur l'état du système. Le nouvel état obtenu suite au déclenchement de l'événement doit préserver l'invariant. Un événement particulier appelé Initialisation doit établir l'invariant. De plus, une machine Event-B peut comporter des théorèmes qui devraient êtres prouvés. Un contexte Event-B comporte les paramètres du système à modéliser : ensembles et constantes. Les propriétés de ces paramètres sont formalisées par des axiomes et des théorèmes à prouver.

Il est à noter qu'un modèle peut contenir uniquement des contextes (c'est un modèle qui représente une structure purement mathématique avec les constantes, les axiomes et les théorèmes), ou bien uniquement des machines (c'est un modèle non paramétré) ou bien les deux ensembles (c'est un modèle paramétré par les contextes). Pour ce troisième type de modèle, il existe un ensemble de relations entre les machines et les contextes, comme représenté dans la **figure 2.1**.

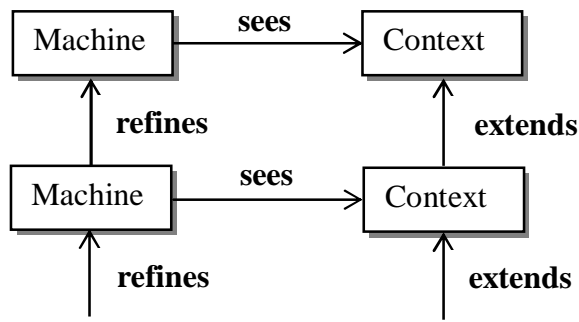

**Figure 2.1 :** *Les relations entre les contextes et les machines*

Les explications liées à la **figure 2.1** couvrent :

- Une machine peut voir explicitement plusieurs contextes (ou pas du tout).
- Un contexte peut étendre explicitement plusieurs contextes (ou pas du tout).
- La notion d'extension de contexte est transitive: un contexte C1 qui étend explicitement un contexte C2, étend implicitement tous les contextes étendus par C2.
- Lorsqu'un contexte C1 étend un contexte C2, alors C1 peut utiliser les ensembles et les constantes de C2.
- Une machine voit implicitement tous les contextes étendus par un contexte qu'elle voit explicitement.
- Quand une machine M voit un contexte C, elle pourra utiliser les ensembles et les constantes de C.
- Il n y'a pas de cycle dans les relations « refines » et « extends ».
- Une machine raffine au plus une autre machine.
- La relation de raffinement « refines » est transitive.

### 2.2.2.1 Contexte

Un contexte « *Context* » permet de spécifier des données statiques. Il se compose d'ensembles, de constantes avec leurs axiomes et éventuellement des théorèmes.
La structure du contexte est constituée d'un ensemble de clauses introduites par des mots clés comme représentée dans la **figure 2.2**.

```
CONTEXT
    Context_identifier
EXTENDS  *
    Context_identifier_list
SETS  *
    set_identifier 1
    ..
    set_identifier m
CONSTANTS  *
    constant_identifier 1
    ..
    constant_identifier n
AXIOMS  *
    label_1: <predicate_1>
    ..
    label_n: <predicate_n>
THEOREMS  *
    label_1: <predicate_1>
    ..
    label_p: <predicate_p>
END
```

**Figure 2.2 :** *Structure d'un contexte*

-Les champs avec " * " peuvent être vides

Le contenu de chaque clause peut être décrit comme suit :

- **CONTEXT** : permet de définir le nom du contexte. Il doit être distinct de tous les autres noms qui existent dans le modèle.
- **EXTENDS** : contient la liste des contextes hérités.
- **SETS** : regroupe la liste des ensembles qui ne sont pas vides et deux à deux disjoints s'ils sont de même type.
- **CONSTANTS** : définit la liste des constantes introduites dans le contexte.
- **AXIOMS** : définit la liste des prédicats que les constantes doivent respecter.
- **THEOREMS** : définit les théorèmes qui vont être prouvés dans le contexte.

Les axiomes et théorèmes portent des étiquettes (labels). Celles-ci peuvent être personnalisées.

### 2.2.2.2    Machine abstraite

En Event-B, on distingue deux types de machine : machine *abstraite* introduite dans le modèle le plus abstrait et machine *de raffinement* (ou raffinée). Une machine contient essentiellement des variables et des évènements. Les variables sont données dans la clause *VARIABLES* et initialisées dans la clause *Initialisation*. La clause *Invariant* définit l'espace d'état des variables et les propriétés du système [41].

La structure d'une machine ressemble à la structure du contexte. Elle admet un ensemble de clauses introduites par des mots clés. La **figure 2.3** montre la structure d'une machine.

```
MACHINE
    Machine_identifier
REFINES  *
    Machine_identifier
SEES  *
    context_identifier_list
VARIABLES
    Variable_identifier_list
INVARIANTS
    label : <predicate>
    .
    .
THEOREMS  *
    label : <predicate>
    .
    .
VARIANT  *
    <variant>
EVENTS
    <event_list>
END
```

**Figure 2.3 :** *Structure d'une machine*

-Les champs avec " * " peuvent être vides

Voici la description du contenu de chaque clause :

- **MACHINE** : donne un identifiant à cette machine. Un tel identifiant doit être différent des identifiants de tous les autres composants du même modèle.

- **REFINES** : représente l'identificateur de la machine (si elle existe) à partir de laquelle raffine la machine actuelle. Une machine raffine au plus une autre machine.

- **SEES :** contient (s'il existe) la liste des contextes vus explicitement par cette machine. Par conséquent, elle peut utiliser les ensembles porteurs (carrier sets) et les constantes vus explicitement ou implicitement par cette machine (par la relation *extends*).

- **VARIABLES :** définit la liste des variables introduites dans la machine.

- **INVARIANTS :** sert au typage des variables et à décrire les contraintes (sous forme de prédicats) qu'elles doivent satisfaire ou respecter en permanence. Un invariant peut être :

    - Invariant *propre* aux variables de la machine courante

    - Invariant de *collage* reliant les variables de la machine concrète avec celles de la machine abstraite (raffinée)

- **THEOREMS :** liste les différents théorèmes qui doivent être prouvés dans cette machine. Pour y parvenir, on utilise comme hypothèses les théorèmes précédents définis avant ce théorème ainsi que les axiomes et les théorèmes des contextes vus explicitement ou implicitement par cette machine.

- **VARIANT :** c'est une clause qui apparaît dans une machine raffinée et qui contient des événements convergents. Cette clause contient une expression de type entier naturel qui décroît par n'importe quel événement convergent ou de type ensemble fini dont sa cardinalité décroît.

- **EVENTS :** détaille les différentes transitions (événements) de la machine.

Dans la suite, nous présentons les constituants d'un événement Event-B.

### 2.2.2.3  Evénement

Tout événement en Event-B modélise une transition discrète et peut être défini par une relation *"avant-après"* notée *BA (x, x')*, où *x* et *x'* désignent respectivement la valeur des variables avant et après l'exécution des actions associées à l'événement. Cette relation varie d'une forme d'événement à une autre [41].

La forme générale d'un événement est représentée par la **figure 2.4** :

```
event_identifier ≙
   STATUS
   {ordinary, convergent, anticipated}
   REFINES  *
      event_identifier
   ANY  *
      parameter_identifier_1
      ..
      parameter_identifier_n
   WHERE  *
      label : <predicate>
      .
      .
   WITH  *
      label : < witness >
      .
      .
   THEN  *
      label : < action >
      .
      .
END
```

**Figure 2.4 :** *Structure d'un événement*

-Les champs avec " * " peuvent être vides

- **STATUS :** décrit l'état d'un événement. Un événement peut être :

    - ordinary,

    - convergent: il doit décrémenter le variant,

    - anticipated: va être convergent ultérieurement dans un raffinement.

- **REFINES** : liste le(s) événement(s) abstraits que l'événement actuel raffine (s'il existe).

- **ANY :** énumère la liste des paramètres de l'événement.

- **WHERE :** contient les différents gardes de l'événement. Ces gardes sont des conditions nécessaires pour déclencher l'événement. Il faut noter que si la clause « any » est omise le mot clé « where » est remplacé par « when ».

- **WITH** : lorsqu'un paramètre dans un événement abstrait disparait dans la version concrète de cet événement, il est indispensable de définir un témoin sur l'existence de ce paramètre : c'est ce qu'on appelle « witness ».

- **THEN :** décrit la liste des actions de l'événement.

Il est à noter que chaque événement est composé d'une ou plusieurs actions dites encore substitution. Nous expliquons, dans ce qui suit, la notion d'action ainsi que ses différentes formes.

### 2.2.2.4    Substitution généralisée

L'action d'un événement peut être soit *déterministe* soit *non déterministe* [41].

- Une action déterministe est de la forme :

> « liste des identificateurs des variables » : = « prédicat avant-après »

En effet, elle se compose d'une liste de variables, suivie du signe « := », suivi d'une liste d'expressions.

- Pour les actions non déterministes, il existe deux formes possibles :

> **(1)** « liste des identificateurs des variables » :| « prédicat avant-après »

> **(2)** « identificateur d'une variable » :$\in$ « expression ensembliste »

La première forme **(1)** correspond à une liste de variables et un prédicat « *avant-après* » séparé par le signe « :| ». La deuxième **(2)** correspond à une variable suivie du signe « :$\in$ », suivie d'une expression ensembliste [32].

### 2.2.2.5    Exemple

La **figure 2.5** montre un exemple illustratif de contexte et de machine notés respectivement **doors_ctx1** et **doors_0** du modèle initial de l'application contrôleur d'accès aux bâtiments destiné à contrôler l'accès aux bâtiments d'un espace de travail. Ce système étant décrit dans [3] et [22]. La machine **doors_0** voit le contexte **doors_ctx1**. En outre, elle admet une variable *sit* désignant la situation d'une personne par rapport à une location et un événement *pass* modélisant l'action de passage d'une location à une autre.

```
CONTEXT
    doors_ctx1
SETS
    P    // Person
    L    // Location
CONSTANTS
    aut    // autorisation
    outside    // à l'extérieur
AXIOMS
    axm_1: aut ∈ P ↔ L    // autorisation de passage des personnes aux bâtiments
    axm_2: outside ∈ L    // l'extérieur est considéré comme une localisation
    axm_3: P×{outside} ⊆ aut    // chacun est autorisé d'être à l'extérieur
    axm_4: ∃l·l∈L\{outside} ∧ P×{l}⊆aut    // l'autorisation d'être dans un bâtiment
END
```

```
MACHINE
    doors_0
SEES
    doors_ctx1
VARIABLES
    sit
INVARIANTS
    inv1 : sit ∈ P → L
    inv2 : sit ⊆ aut
EVENTS
    INITIALISATION ≙
        STATUS
            ordinary
        BEGIN
            act1 : sit ≔ P×{outside}
    END

    pass ≙
        STATUS
            ordinary
        ANY
            p
            l
        WHERE
            grd1 : p ↦ l ∈ aut    // pour passer on doit avoir une autorisation
            grd2 : sit(p) ≠ l    // p ne doit pas être dans l
        THEN
            act1 : sit(p) ≔ l
    END
END
```

**Figure 2.5 :** *Exemple de contexte et de machine*

## 2.2.3 Obligations de preuves

Afin de garantir la correction de notre modèle, il est indispensable de le prouver. Pour y parvenir, un outil de la plateforme RODIN appelé ***générateur d'obligations de preuve*** génère automatiquement des *obligations de preuve* [37]. Une obligation de preuve définit ce que doit être prouvé pour un modèle. Il s'agit d'un prédicat dont on doit fournir une démonstration pour vérifier un critère de correction sur le modèle.

L'outil cité ci-dessus vérifie statiquement les contextes et les machines et génère des *séquents*. « Un séquent est un nom générique pour quelque chose qu'on veut prouver » [3]. Il est de la forme **H ⊢ G** : 'le but G est à démontrer en partant de l'ensemble H des hypothèses'.

Les règles des obligations de preuve sont au nombre de onze (11), et pour chacune le générateur d'obligations de preuve génère une forme spécifique du séquent.

Etant donné un élément de modélisation; un événement **evt**, un axiome **axm**, un théorème **thm**, un invariant **inv**, une garde **grd**, une action **act**, un variant ou une witness **x**. les règles d'obligation de preuve pouvant être générées pour ces éléments sont les suivantes :

- ***INV*** : règle de préservation de l'invariant qui assure que chaque invariant dans une machine donnée est préservé par tous les événements. Elle est de la forme : *"evt/inv/INV"*.
- ***FIS*** : se rassurer qu'une action non déterministe est faisable. La forme est: *"evt/act/FIS"*.
- ***GRD*** : renforcement des gardes abstraits ; les gardes des événements concrets sont plus fortes que celles des abstractions. Elle est de la forme : *"evt/grd/GRD"*.
- ***MRG*** : la garde d'un événement concret qui fusionne deux événements abstraits est plus forte que la disjonction (ou logique) des gardes de ces deux événements abstraits. Le nom de cette règle est: *"evt/MRG"*.
- ***SIM*** : chaque action dans un événement abstrait est correctement simulée dans le raffinement correspondant. Autrement dit, l'exécution d'un événement concret n'est pas en contradiction avec son abstraction. Le nom : *"evt/act/SIM"*.
- ***NAT*** : sous une condition que les gardes d'un événement convergent ou anticipé sont vérifiées, le variant numérique proposé est un *entier naturel*. Son nom est : *"evt/NAT"*.
- ***FIN*** : dans une condition où les gardes d'un événement convergent ou anticipé sont vérifiées, l'ensemble variant proposé est un ensemble *fini*. Son nom est : *"evt/FIN"*.

- *VAR :* chaque événement convergent diminue le variant numérique proposé ou l'ensemble variant proposé. De plus, chaque événement anticipé n'augmente pas le variant numérique proposé ou l'ensemble variant proposé. Son nom est : *"evt/VAR "*.
- *WFIS :* chaque témoin (witness) proposé dans la clause *WITH* (si elle existe) d'un événement concret existe vraiment. Son nom est : *"evt/x/WFIS "*.
- *THM :* un théorème écrit dans une machine ou dans un contexte est vraiment prouvable. Le nom de telle règle est : *"thm/THM "*.
- *WD :* règle de bonne définition des axiomes, théorèmes, invariants, gardes, actions, variant et witness. Selon la nature de l'élément, les noms pour cette règle sont : *"axm/WD "*, *"thm/WD "*, *"inv/WD "*, *"grd/WD "*, *"act/WD "*, *"VWD "* ou *"evt/x/WWD"*.

### 2.2.4 Raffinement

#### 2.2.4.1 Généralités sur la relation de raffinement

Le raffinement [44] est le processus de construction d'un modèle progressivement en le rendant de plus en plus précis. Ce qui aboutit à un modèle très proche de la réalité. Il consiste à construire une séquence ordonnée de modèles où chacun est considéré comme un raffinement d'un autre modèle précédent de la séquence.

Cette technique établit un contrat entre ces deux modèles (partenaires) : un composant dit **abstrait** et un composant dit **concret**. Elle est potentiellement utilisée dans plusieurs niveaux de développement des logiciels :

- *Dans la phase de spécification :* c'est un moyen d'ajout de détails du problème dans le développement formel. Ceci est traduit en Event-B par l'adjonction de nouvelles variables et de nouveaux événements.

- *Dans la phase de conception :* raffinement du diagramme de classes initial, formé par des classes d'analyse (classes métier), par ajout des classes de conception et puis des classes d'implémentation.

- *Dans la phase d'implémentation :* développement progressif des programmes [44] : raffinement des instructions et des données.

### 2.2.4.2    Obligations de preuves de raffinement

Dans une machine raffinée, le générateur d'obligations de preuves génère automatiquement des obligations de preuves relatives au raffinement. Parmi ces obligations de preuves, on cite :

- **INV :** utilisée pour démontrer la préservation de l'invariant du raffinement. Pour un événement *evt* qui raffine un événement *evt0*, la règle d'obligation de preuves **INV** relative à la préservation de l'invariant pour l'événement evt est de la forme :

| | |
|---|---|
| Axiomes et théorèmes<br>Invariants et théorèmes abstraits<br>Invariants et théorèmes concrets<br>Gardes de l'événement concret<br>prédicat témoin pour les variables<br>Prédicat concret avant-après<br>⊢<br>Invariant spécifique modifié | **evt/inv/INV** |

- **GRD** : permet d'assurer que la garde concrète d'un événement raffiné est renforcée par rapport à celui de l'abstrait. Pour un événement *evt* concret et une garde *grd* abstraite la forme de cette obligation de preuve est la suivante :

| | |
|---|---|
| Axiomes et théorèmes<br>Invariants et théorèmes abstraits<br>Invariants et théorèmes concrets<br>Gardes de l'événement concret<br>prédicat témoin<br>⊢<br>Garde spécifique de l'événement abstrait | **evt/grd/GRD** |

- **SIM :** assure que pour chaque action appartenant à un événement abstrait est correctement simulé dans le raffinement correspondant. C'est-à-dire quelle assure qu'un événement concret n'est pas opposé à l'événement abstrait qui lui correspond. Soit *evt0* un événement, *evt* son raffinement et *act* une action abstraite, cette obligation de preuve est de la forme :

| | |
|---|---|
| Axiomes et théorèmes<br>Invariants et théorèmes abstraits<br>Invariants et théorèmes concrets<br>Gardes de l'événement concret<br>prédicat témoin des paramètres | **evt/act/SIM** |

| prédicat témoin des variables | |
| Prédicat concret avant-après | |
| ⊢ | |
| Prédicat abstrait avant-après | |

- **MRG :** assure qu'une garde d'un événement concret combinée par deux événements abstraits est plus renforcée que celle de la disjonction des gardes des deux événements. Soient *evt01* et *evt02* deux évènements abstraits admettant les mêmes paramètres et actions et dont *evt* est l'événement concret combiné, cette obligation de preuve est de la forme :

| Axiomes et théorèmes | |
| Invariants et théorèmes abstraits | |
| Gardes de l'événement concret | **evt/MRG** |
| ⊢ | |
| Disjonction des gardes abstraites | |

### 2.2.4.3 Types de raffinement

On peut distinguer deux types de raffinements [3] : *le raffinement horizontal* et *le raffinement vertical*.

- *Le raffinement horizontal ou superposition:* lorsqu'on a un système complexe contenant plusieurs composants et avec des transitions discrètes, on ne peut pas le spécifier d'un seul coup. On a intérêt à le spécifier progressivement pas-à-pas en commençant par un modèle très abstrait et en introduisant à chaque pas des détails issus du cahier des charges réécris conformément à la stratégie de raffinement adoptée.

- *Le raffinement vertical :* on l'adopte lorsque toutes les étapes de raffinement horizontal sont achevées. Il est guidé par des décisions de conception ou d'implémentation. Le raffinement vertical a pour objectif d'apporter une solution pas-à-pas à la spécification produite par le raffinement horizontal.

### *2.2.4.4 Exemple*

Soit l'événement abstrait **check_out** d'une application de gestion de réservations dans un hôtel (voir chapitre 3).

```
check_out  ≜    //  Annulation des réservations
   STATUS
    ordinary
   ANY
      g
      r
   WHERE
      grd1 : r↦g∈owns   // r doit être réservée par g
   THEN
      act1 : owns≔owns\{r↦g}   // la chambre r devient non réservée
   END
```

Son raffinement appelé **check_out1** (1 pour dire premier raffinement) obéit au protocole cité ci-dessus. En fait, il est augmenté par un seul paramètre c, de plus ses gardes sont renforcées et de même pour ses substitutions (actions).

Les éléments ajoutés sont colorés en bleu.

```
check_out1  ≜    //  Annulation des réservations
   STATUS
    ordinary
   REFINES
      check_out
   ANY
      g
      r
      c
   WHERE
      grd1 : r↦g∈owns   // r doit être réservée par g
      grd2 : c↦g∈cards   // g doit être le propriétaire de c
   THEN
      act1 : owns≔owns\{r↦g}   // la chambre r devient non réservée
      act2 : cards≔cards\{c↦g}    // carte retirée du client
   END
```

En outre, une seule obligation de preuve relative à ce raffinement est déchargée automatiquement : *check_out1/inv1_1/INV*.

### 2.2.5 Propriété de vivacité

Une propriété de sûreté stipule que rien de mauvais n'arrive. En Event-B, le générateur d'Obligations de Preuves (OP) produit des OP afin de vérifier les propriétés de sûreté. Ces OP concernent la préservation d'invariants et la correction de raffinement (voir **2.2.3**). La démonstration de ces OP en se servant du prouveur d'Event-B garantit la cohérence des modèles Event-B.

Une propriété de vivacité stipule que quelque chose de bien arrive. Le générateur d'obligations de preuves produit des OP liées à la convergence des nouveaux événements introduits dans une étape de raffinement (evt/VAR voir **2.2.3**). Ces OP couvrent la propriété de vivacité suivante : les nouveaux événements ne se déclenchent qu'un nombre fini de fois. Ceci permet aux (anciens) événements abstraits de prendre la main.

En Event-B, le non-blocage du système peut être vérifié par la démonstration d'un théorème manuel dont l'expression est la disjonction des gardes des événements. Un tel théorème est souvent décrit par une grosse formule difficile à démontrer.

Les propriétés de vivacité spécifiques qui traduisent des comportements dynamiques peuvent être décrites par une logique temporelle et vérifiées par le model-checker ProB [21].

### 2.2.6 Plateforme RODIN

La plateforme RODIN [33] [34] [43] est un environnement dédié à Event-B [16]. Hormis la structure d'accueil offerte par RODIN, ce dernier intègre les composants logiciels suivants :
- Un éditeur dirigé par la syntaxe pour saisir les modèles Event-B.
- Un analyseur lexico-syntaxique d'Event-B.
- Un vérificateur de typage d'Event-B.
- Un générateur d'obligations de preuve selon les règles exposées dans la section **2.2.3**.
- Un prouveur automatique et interactif assez puissant.

La **figure 2.6** présente une vue macroscopique de la plateforme RODIN et la **table 2.1** fournit la signification des différents éléments de cette vue macroscopique.

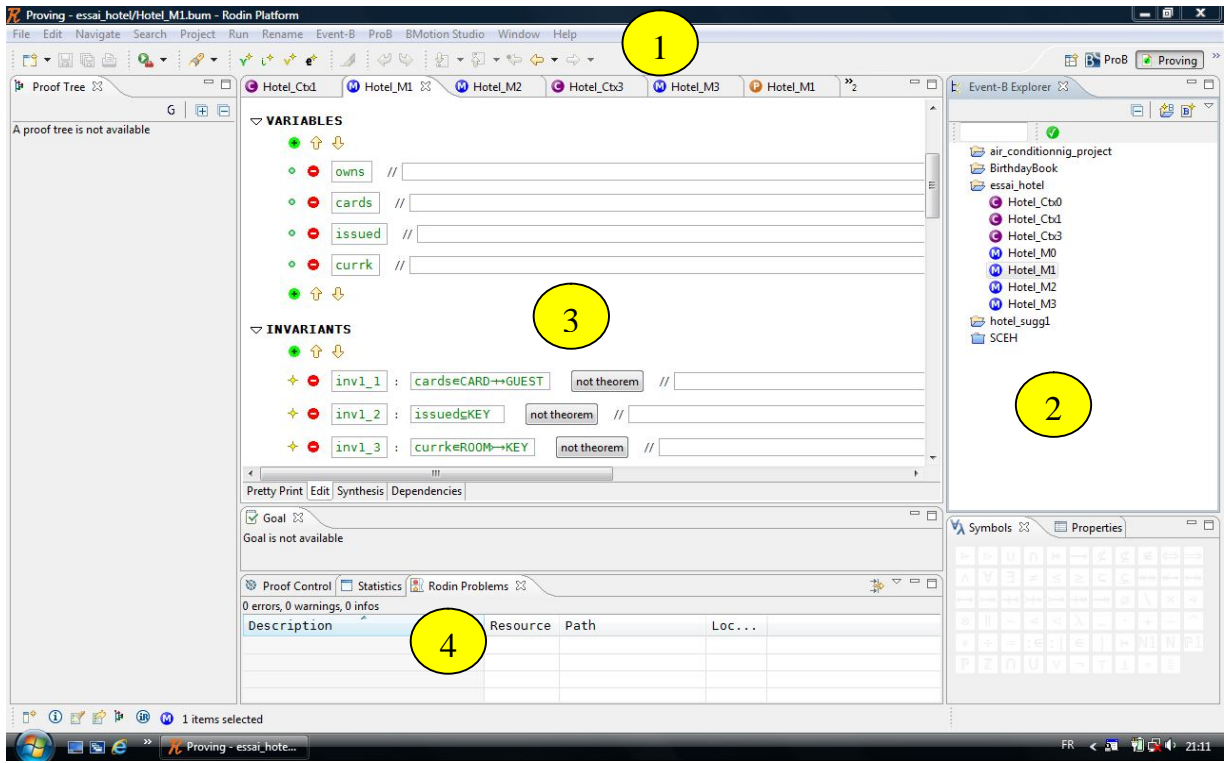
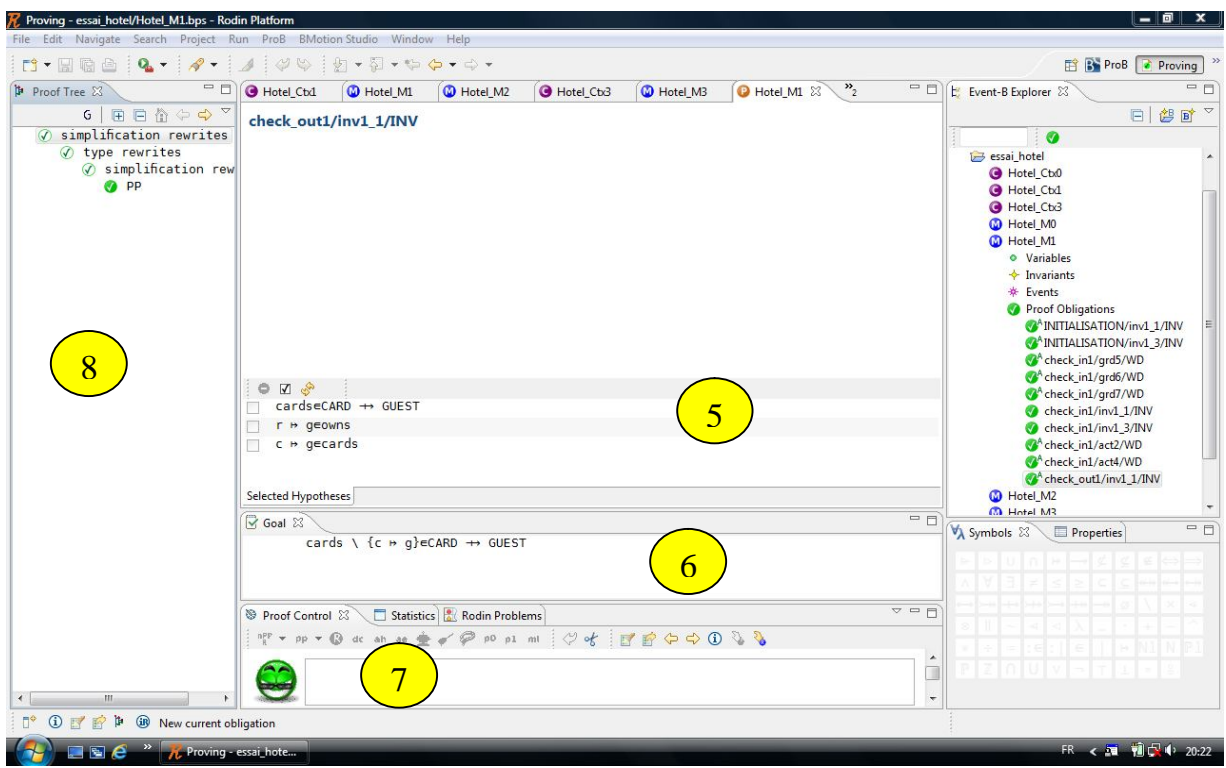

**Figure 2.6 :** *Une vue macroscopique de la plateforme RODIN*

| 1 | Barre d'outils |
|---|---|
| 2 | Explorateur de projets |
| 3 | Zone de saisie |
| 4 | Problèmes rencontrés |
| 5 | Hypothèses sélectionnées |
| 6 | But à atteindre |
| 7 | Contrôle de preuves |
| 8 | Arbre de preuve |

**Table 2.1 :** *Structure d'accueil de la plateforme RODIN*

## 2.3  La bibliothèque EM-OCL

Le langage OCL [42] fait partie intégrante de la dernière version d'UML [29]. Il est utilisé au niveau modélisation afin de spécifier formellement les contraintes associées aux modèles UML tout en s'inspirant de la conception par contrat (Design by contract) introduite par Bertrand Meyer [25]. En outre, il est utilisé au niveau méta-modélisation afin de spécifier la sémantique statique formelle des langages de modélisation comme UML et les profils UML. Cependant, OCL montre des insuffisances dans divers domaines. En effet, la mise en œuvre de la technique de raffinement [44], qui constitue la pierre angulaire des méthodes formelles comme B [4] et Event-B [3], exige des moyens de description des données -partie statique de l'application- adéquats tels que : ensemble, relation, fonction totale fonction partielle, fonction surjective, fonction injective et fonction bijective. De plus, la validation des diagrammes de classes basée sur des vues issues du monde réel modélisées par des diagrammes d'objets nécessitent des contraintes OCL attachées aux diagrammes de classes. De telles contraintes traduisent des propriétés invariantes globales pratiquement inexprimables en OCL. En effet, celui-ci permet la description des propriétés invariantes relatives à un contexte (*Classifier*). Enfin, le langage OCL offre des possibilités favorisant son utilisation en tant que langage de requêtes telles que : *collect*, *select* et *reject*. Mais les requêtes écrites en OCL doivent être liées à l'objet courant *self*. Ceci entraîne des difficultés relatives à l'écriture de telles requêtes.

Afin d'ouvrir OCL sur les domaines cités ci-dessus (raffinement, validation de diagrammes de classes, langage des requêtes), notre équipe a proposé une extension mathématique au langage OCL appelée **EM-OCL** permettant de représenter et manipuler les concepts mathématiques suivants : couple, relation binaire et fonction. Ces concepts sont modélisés par des classes génériques ayant des propriétés invariantes bien définies en EM-OCL en combinant généricité et plusieurs genres d'héritage [29] (héritage d'extension, héritage de sous type, héritage de restriction, héritage de variation de type et héritage de variation fonctionnelle). Ces classes sont intégrées d'une façon judicieuse dans la bibliothèque des classes OCL. Les opérations offertes par ces classes sont définies formellement en utilisant une spécification pré/post supportée par OCL et par conséquent par EM-OCL.

Notre extension EM-OCL propose des augmentations liées au package Types d'OCL afin de typer (ou méta-modéliser) les concepts mathématiques déjà cités. En outre, elle propose des enrichissements liés à la bibliothèque standard d'OCL dans le but de définir d'une façon formelle –en utilisant OCL lui-même – les opérations applicables sur les nouveaux types introduits dans EM.

### 2.3.1 Les augmentations liées au package Types d'OCL

Le langage OCL est fortement typé : il est indispensable de donner un type à chaque élément utilisé. Pour y parvenir, EM-OCL apporte de nouveaux types modélisant les concepts mathématiques couple, relation binaire et fonction.

La **figure 2.7** montre les méta-classes (colorées) ajoutées au package Types d'OCL. Toutes les méta-classes -sauf PairType et SequenceRefType- héritent de la méta-classe SetType.

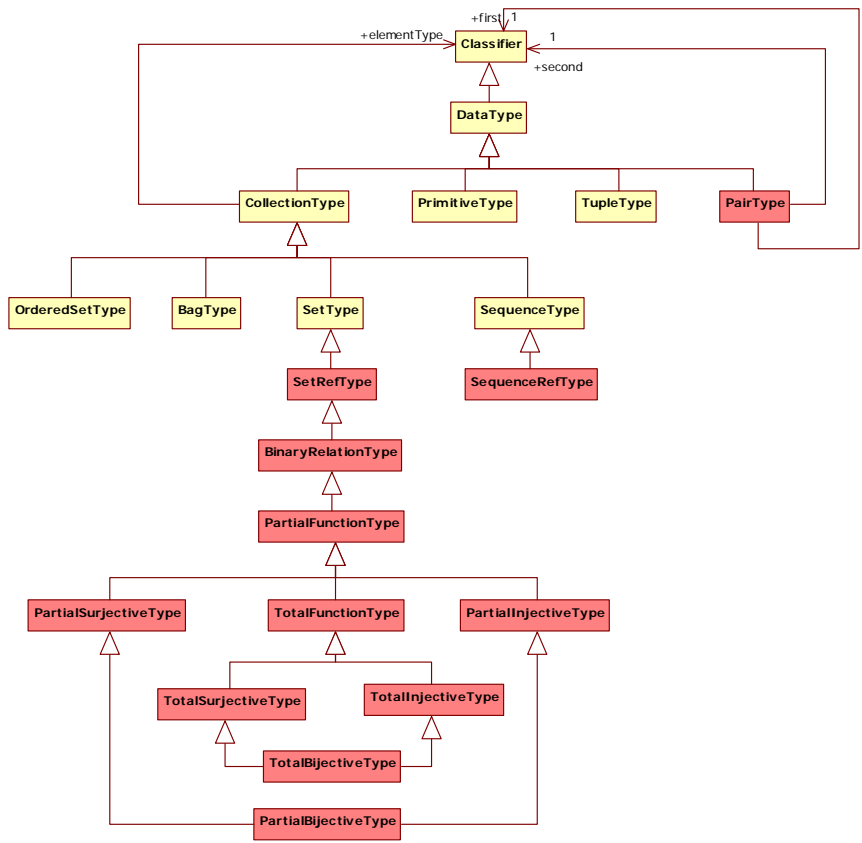

**Figure 2.7 :** *Une partie du méta-modèle des types EM-OCL*

La méta-classe *PairType* modélise le concept d'une paire ordonnée offert par EM-OCL. Elle descend de la méta-classe abstraite *DataType*. Elle se distingue de la méta-classe *TupleType* par le fait que celle-ci modélise le concept d'une structure comportant plusieurs champs dont l'ordre n'est pas significatif à l'instar de type *struct* de *C* ou *record* de Pascal ou Ada. Les deux méta-associations *first* et *second* entre *PairType* et *Classifier* modélisent la nature de deux éléments formant la paire ordonnée.

Les méta-classes *BinaryRelationType, PartialFunctionType, TotalFunctionType, PartialInjectiveType, TotalInjectiveType, PartialSurjectiveType, TotalSurjectiveType, PartialBijectionType* et *TotalBijectionType* représentent respectivement les concepts mathématiques relation binaire, fonction partielle, fonction totale, fonction injective partielle, fonction injective totale, fonction surjective partielle, fonction bijective partielle et fonction bijective totale.

Les deux méta-classes *SequenceRefType* et *SetRefType* apportent des nouvelles opérations aux deux méta-classes *SequenceType* et *SetType* d'OCL.

Les liens entre les méta-classes apportées par EMF-OCL sont traduits par des relations d'héritage simples et multiples inspirées de la définition mathématique des notions : couple, relation binaire et fonction [4].

A l'instar d'OCL, le package Types de la bibliothèque EM-OCL propose des règles de conformité des nouveaux types supportés. De telles règles sont définies comme des propriétés invariantes établies dans le contexte de la méta-classe du type concerné. Egalement, cette dernière (EM-OCL) propose des règles de bonne utilisation (well-formedness rules) pour ses nouveaux types. Ces règles sont décrites dans [40].

### 2.3.2  Les augmentations liées à la bibliothèque standard d'OCL

La **figure 2.8** montre les notions mathématiques intégrées dans EM-OCL sous forme de classes génériques organisées dans une hiérarchie.

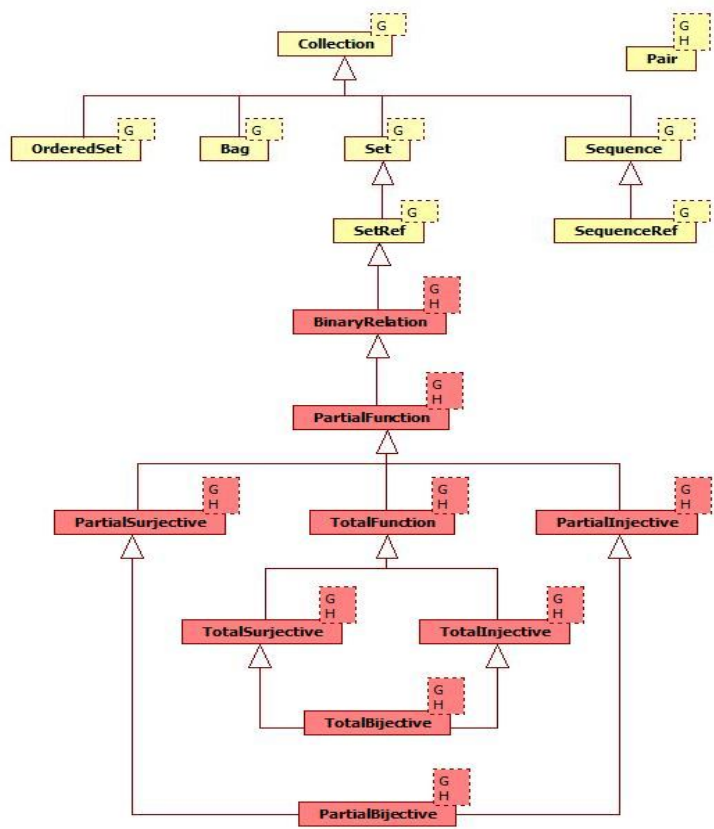

**Figure 2.8 :** *Graphe des classes de la bibliothèque EM ajoutée à OCL*

Afin d'intégrer dans le langage OCL les concepts mathématiques paire, relation et fonction, nous avons suivi un processus guidé par les types abstraits de données. Ainsi, toutes les classes apportées par EM-OCL modélisent des notions bien définies c'est-à-dire dotées des opérations formellement spécifiées (en utilisant les deux clauses *pre* et *post* d'OCL) et des propriétés invariantes (en utilisant la clause *inv* d'OCL). Sachant que les deux classes d'OCL modélisant les types collectifs (*Bag, Set, Collection, OrderedSet* et *Sequence*) ne sont pas équipées des propriétés invariantes.

Nous avons combiné avec profit la généricité et l'héritage (simple, multiple et répété) afin de découvrir les relations conceptuelles pertinentes fortes entre les classes formant la bibliothèque EM-OCL.

Les classes ayant un seul paramètre générique formel sont paramétrées sur la nature des éléments des conteneurs qui descendent directement ou indirectement de *Collection*. Les classes qui descendent directement ou indirectement de *BinaryRelation* possèdent deux paramètres génériques formels *G* et *H*. Ces deux paramètres désignent respectivement le type de l'ensemble de départ et d'arrivée. Enfin, la classe *Pair* modélise un couple au sens mathématique. Elle admet deux paramètres génériques formels *G* et *H* désignant respectivement la nature du premier et second élément du couple.

Les hiérarchies d'héritage proposées par la bibliothèque EM-OCL comporte plusieurs genres d'héritage [42]: héritage d'extension, héritage de sous type, héritage de restriction, héritage de variation de type et héritage de variation fonctionnelle.

Les opérations offertes par notre bibliothèque EM sont décrites dans [40] et [8].

En conclusion, la bibliothèque EM-OCL, proposée par notre équipe, permet de représenter et manipuler les concepts mathématiques couple, fonction totale, fonction partielle, fonction injective, fonction surjective, fonction bijective et relation binaire dans un cadre OO (Orienté Objet). Ces derniers sont modélisés par des classes génériques. Ces classes sont intégrées d'une façon judicieuse dans la bibliothèque de classes OCL. Les opérations offertes par ces classes sont définies formellement en utilisant une spécification pré/post supportée par OCL et par conséquent par EM-OCL. Les propriétés invariantes des classes de la bibliothèque EM ont été formulées en utilisant une approche orientée invariant. Des utilisations potentielles de notre langage EM-OCL ont été exhibées sur des exemples bien ciblés, dans divers domaines : développement des diagrammes de classes en utilisant la technique de raffinement, validation de diagrammes de classes et utilisation d'OCL en tant que langage de requêtes [8] [40].

**Dans ce travail, nous élargissons le champ d'action d'EM-OCL en l'utilisant en tant que langage pivot entre Event-B et UML/OCL.**

## 2.4  Conclusion

Dans ce chapitre, nous avons présenté d'une façon rigoureuse les concepts de base de la méthode formelle Event-B. Une telle méthode permet le développement pas-à-pas des logiciels corrects par construction en se servant de la technique de raffinement. De plus, nous avons exposé les principaux composants de la bibliothèque EM-OCL permettant de modéliser les concepts mathématiques paire, fonction et relation.

Dans la suite de ce travail, nous allons utiliser Event-B pour modéliser formellement la spécification de l'application Système de Clés Electroniques pour Hôtels (SCEH) [3] et UML/EM-OCL comme langage pivot entre Event-B et UML/OCL.

# Chapitre 3 : Spécification en Event-B de l'application Système de Clés Electroniques pour Hôtels (SCEH)

## 3.1 Introduction

Dans ce chapitre, nous comptons établir une spécification formelle en Event-B de l'application Système de Clés Electroniques pour Hôtels (SCEH) décrit par J-R Abrial dans [3]. Pour y parvenir, nous allons appliquer les phases en amont de l'approche hybride proposée dans le chapitre 1. Ces phases concernent la *Réécriture* du cahier des charges de l'application SCEH, *Spécification* abstraite et *Raffinement* horizontal. Enfin, nous allons utiliser le model-checker ProB [21] afin d'animer les différents modèles Event-B établis liés à l'application SCEH.

## 3.2 Cahier des charges

### 3.2.1 Présentation informelle

Un hôtel [3] utilisant des clés métalliques risque d'être non sécurisé en termes de droits d'accès aux chambres. En fait, une clé métallique peut avoir une copie et, par conséquent, l'accès à la chambre correspondante peut être possible à tout moment par n'importe quel client précédent. L'utilisation judicieuse d'un système de clés électroniques approprié pourrait garantir l'unicité d'accès aux chambres par leurs clients actuels. Le système souhaité se caractérise par les propriétés suivantes :

1) Chaque porte est équipée d'une serrure électronique mémorisant une clé électronique. Une telle serrure est dotée d'un lecteur de cartes magnétiques.

2) L'enregistrement d'un client est suivi d'une réservation de chambre. Suite à chaque réservation, une carte est livrée au client locataire de la chambre.

3) Une carte magnétique maintient deux clés électroniques : une clé k1 identique à celle mémorisée par la serrure de la porte correspondante et modélisant le client précédent, et une nouvelle clé k2 qui indique le client actuel.

4) A chaque nouvelle carte insérée au fond de la serrure, une vérification et une comparaison de clés s'effectuent : si la clé mémorisée dans la serrure est l'une des clés de la carte, alors, la carte sera acceptée, la clé de la serrure sera remplacée par la nouvelle clé k2 de la carte et la porte s'ouvre, ainsi, le client pourra désormais ré-entrer à sa chambre avec la même carte, autrement (clé non convenable) la carte est erronée, elle sera donc rejetée et la porte est maintenue bloquée. Ceci est illustré par la **figure 3.1**.

5) Chaque chambre est sous la responsabilité d'un personnel ayant lui aussi le droit d'y accéder via une carte.

6) Un client doit utiliser sa chambre au moins une fois afin d'autoriser l'accès à cette chambre par un futur locataire.

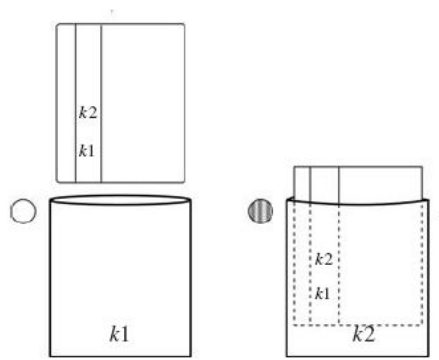

**Figure 3.1 :** *Une serrure de porte et une carte y étant insérée*

### 3.2.2 Restructuration du cahier de charges : texte référentiel

Dans notre application, nous allons nous intéresser uniquement à la rédaction du *texte référentiel* (voir chapitre 1) du cahier de charges cité ci-dessus.

- La fonction principale du système à spécifier est exprimée comme suit :

    Le système doit garantir à chaque client l'accès unique à sa chambre d'hôtel.

| L'accès à une chambre est limité au client qui l'a réservée. | FUN-1 |
|---|---|

- Chaque porte est équipée d'une serrure électronique mémorisant une clé électronique au fond de laquelle on peut insérer une carte magnétique afin d'y accéder.

| Chaque porte est équipée d'une serrure électronique mémorisant une clé électronique et dotée d'un lecteur de cartes magnétiques | ENV-1 |
|---|---|

- Une carte magnétique contient deux clés électroniques.

| Une carte magnétique contient deux clés électroniques distinctes : une première clé k1 modélisant le client précédent et une deuxième clé k2 modélisant le client actuel de la chambre | ENV-2 |
|---|---|

| Les employés de l'hôtel ont le droit d'accéder aux chambres sous leurs responsabilités avec des cartes identiques à celles des clients | FUN-2 |
|---|---|

| La première visite d'un client à sa chambre est traduite par une mise à jour de la clé enregistrée dans la serrure | FUN-3 |
|---|---|

| L'accès aux chambres est contrôlé par des cartes magnétiques | FUN-4 |
|---|---|

## 3.3  Stratégie de raffinement proposée

Avant d'entamer la spécification formelle de notre système, il est indispensable d'identifier la stratégie de raffinement à suivre. Une telle stratégie concerne l'ordre dans lequel nous allons prendre en compte les contraintes du texte référentiel citées précédemment. La stratégie proposée est la suivante :

- *Modèle abstrait initial : Réservation dans l'hôtel*
    On commence avec un modèle abstrait très simple modélisant une application classique de réservation de chambres dans un hôtel. La contrainte ou propriété concernée est **FUN-1**.

- *Premier raffinement : Introduction de la notion de carte magnétique*
    Les contraintes concernées par ce raffinement sont **ENV-2** et **FUN-4**.

- *Deuxième raffinement : Introduction de la notion de serrure électronique*

    Les contraintes concernées par ce raffinement sont **ENV-1** et **FUN-3**.

- *Troisième raffinement : Personnel de l'hôtel*

    La contrainte concernée par ce raffinement est **FUN-2**.

La stratégie de raffinement horizontal permettant à terme d'obtenir une spécification formelle de notre application SCEH est illustrée par la **figure 3.2**.

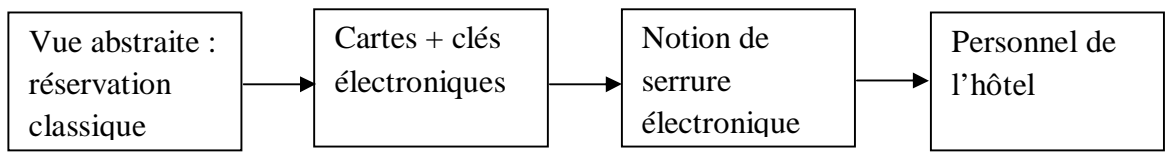

**Figure 3.2 :** *Processus de raffinement horizontal proposé*

## 3.4 Spécification du système SCEH en Event-B

L'architecture générale de la modélisation en Event-B de l'application SCEH est donnée par la **figure 3.3**. Elle comporte une machine abstraite (Hotel_M0), trois machines affinées (Hotel_M1, Hotel_M2 et Hotel_M3) et trois contextes (Hotel_Ctx0, Hotel_Ctx1 et Hotel_Ctx3). Trois types de relations sont utilisées dans cette architecture : la relation **refines** liant soit une machine abstraite à une machine affinée, soit une machine affinée de niveau *i* à une machine affinée de niveau *i+1*, la relation **extends** entre deux contextes et la relation **sees** entre soit une machine abstraite et un contexte, soit entre une machine affinée et un contexte. Sachant que les deux relations **refines** et **extends** sont transitives.

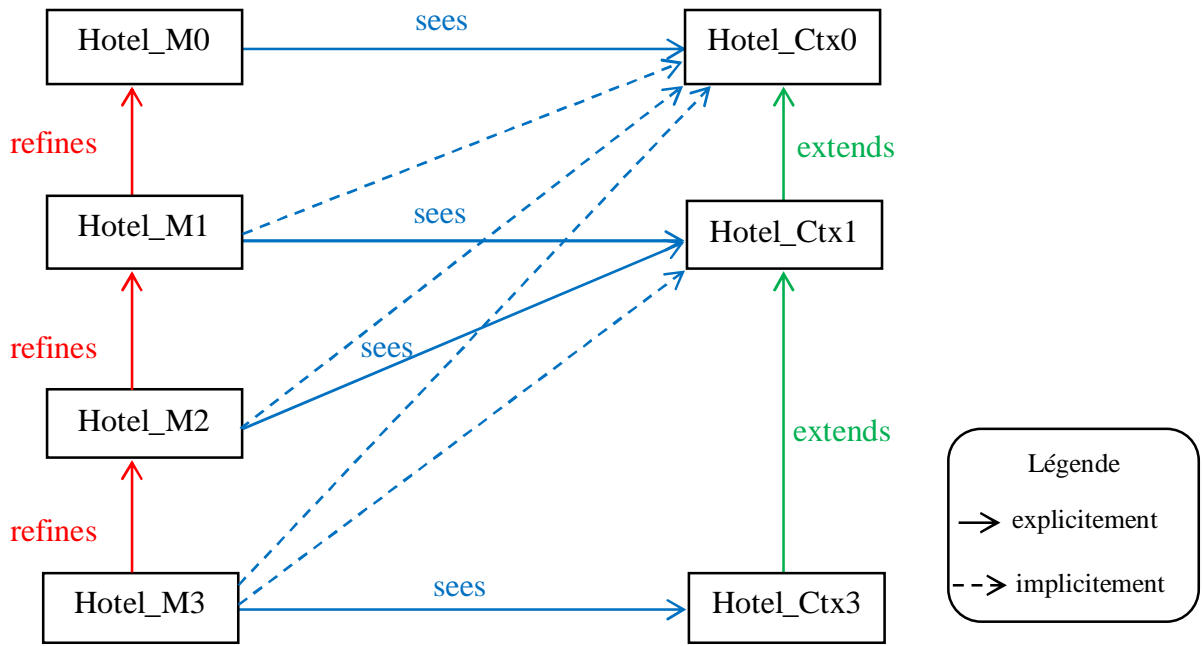

**Figure 3.3 :** *Architecture générale du modèle proposé en Event-B de l'application SCEH*

Afin de modéliser pas-à-pas (step-by-step) l'application SCEH, nous avons utilisé un processus de raffinement horizontal (voir **2.2.4.3**). En outre, nous avons suivi une démarche orientée état afin d'établir les constituants d'une machine. Une telle démarche comporte les étapes suivantes :

- définir les *paramètres* (ensembles porteurs, constantes, axiomes et théorèmes) en les regroupant au sein d'un contexte pouvant étendre un contexte existant.

- définir l'*état* d'une machine par une ou plusieurs variables typées. Lors de cette étape, un soin particulier est accordé aux propriétés invariantes délimitant les états admis vis-à-vis de l'espace d'états potentiel.

- définir les *transitions* autorisées permettant de changer l'état d'une machine. Ceci est exprimé par des événements en Event-B.

### 3.4.1 Modèle initial abstrait: Droits de réservation des chambres limités aux clients

Il s'agit du modèle le plus simple et le plus abstrait de notre spécification. On assimile l'application SCEH à une *application classique de réservation dans un hôtel*.

L'énoncé de la contrainte **FUN-1** à traiter à ce niveau étant le suivant :

| | |
|---|---|
| L'accès à une *chambre* est limité au *client* qui l'a *réservée*. | FUN-1 |

### 3.4.1.1 Les paramètres :

Les paramètres du système sont regroupés au sein d'un *contexte*. Celui-ci comporte deux ensembles abstraits (Sets en Event-B) **GUEST** et **ROOM** modélisant respectivement l'ensemble des clients potentiels et l'ensemble de toutes les chambres de l'hôtel.

Par conséquent, notre système est initialement paramétré sur les ensembles abstraits déclarés comme suit :

**SETS**
  GUEST   *// les clients potentiels*
  ROOM    *// les chambres de l'hôtel*

### 3.4.1.2 Etat du système

L'état du système est décrit par une *variable* appelée *owns* modélisant la distribution des droits de réservation entre clients et typée par l'invariant *inv0_1*. Ce dernier indique le fait qu'une chambre peut être associée à au plus un client : *fonction partielle* $\nrightarrow$. Tandis qu'un client peut effectuer plusieurs réservations.

**VARIABLES**
  owns   *// les propriétaires des chambres*

**INVARIANTS**
  inv0_1  : owns∈ROOM$\nrightarrow$GUEST

Ainsi, on peut définir l'état initial de notre système. Un tel état est représenté dans l'événement ***INITIALISATION*** par une substitution (action) notée *act1* modélisant le fait qu'on a commencé avec un état où on n'a enregistré *aucune réservation*.

```
INITIALISATION ≙    // Initialisation du système
  STATUS
    ordinary
  BEGIN
    act1: owns ≔ ∅    // aucune réservation enregistrée
  END
```

L'obligation de preuve relative à cet événement a été déchargée automatiquement par le prouveur de la plateforme **RODIN**. Elle est de la forme *INITIALISATION/inv0_1/INV* et signifie que l'événement d'initialisation préserve l'invariant *inv0_1*.

### 3.4.1.3 Les transitions

Ce système comporte deux *transitions* dites encore *événements* (autres que l'événement INITIALISATION) appelés *check_in* et *check_out* et modélisant respectivement les actions d'**enregistrement** et de **terminaison** d'une réservation.

- *L'événement check_in :*

Cet événement nécessite, à ce niveau, deux paramètres : un client *g* et une chambre *r*. Ces derniers doivent respecter trois gardes notées ***grd1***, ***grd2*** et ***grd3*** dont les deux premières sont destinées au typage des paramètres et la dernière exprime une contrainte sur l'état du couple ordonné (r, g) par rapport à la fonction *owns* : r doit être non encore réservée. Ci-dessous la spécification en Event-B de cet événement :

```
check_in ≙    // Gestion des réservations
  STATUS
    ordinary
  ANY
    g
    r
  WHERE
    grd1: g∈GUEST    // g est un client
    grd2: r∈ROOM    // r est une chambre
    grd3 : r∉dom(owns)    // la chambre r n'est pas réservée
  THEN
    act1: owns(r)≔g    // le client g devient le propriétaire de la chambre r
  END
```

Un tel événement va agir sur la variable *owns* pour la modifier. Cela peut être illustré ci-dessous par une représentation graphique de la fonction owns avant et après une demande de réservation de la chambre $r_m$ effectuée par le client $g_i$ :

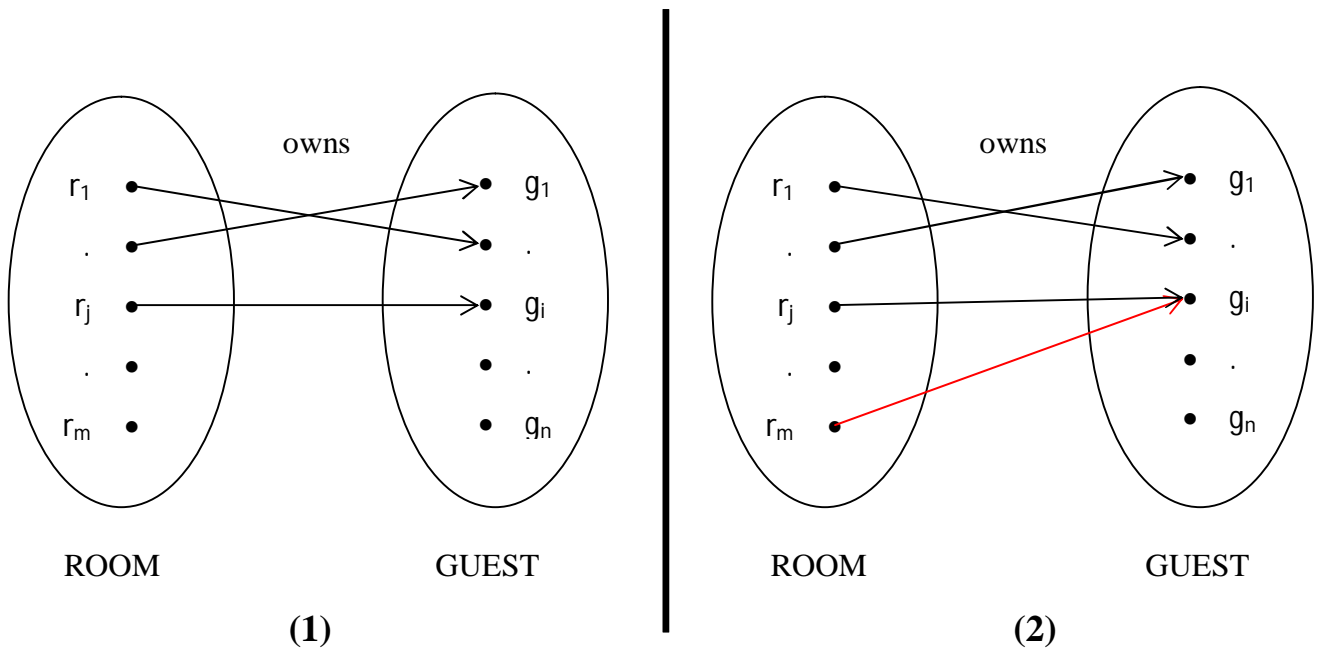

**Figure 3.4 :** *La variable owns en forme digitale avant (1) et après (2) la transition check_in($g_i$, $r_m$)*

- *L'événement check_out :*

    Il concerne la terminaison de réservation. Pour y parvenir, il nécessite deux paramètres : un client g et une chambre r. De plus, la contrainte sur ces paramètres est exprimée via une garde notée *grd1* indiquant le fait que r doit être réservée par g.

```
check_out ≙    // Annulation de réservations
   STATUS
     ordinary
   ANY
      g
      r
   WHERE
      grd1 : r↦g∈owns    // g est le propriétaire de r
   THEN
      act1: owns≔owns\{r↦g}    // la chambre r devient non réservée
   END
```

Les deux obligations de preuve relatives à la préservation de l'invariant *inv0_1* ont été déchargées automatiquement par le prouveur de la plateforme **RODIN**.

### 3.4.1.4  Modèle initial abstrait en Event-B

    A ce niveau d'abstraction, on a proposé deux composants Event-B : Le premier est un contexte appelé **Hotel_Ctx0** (pour dire contexte niveau 0) et le second est une machine **Hotel_M0** (pour dire machine niveau 0) qui voit explicitement **Hotel_Ctx0**.

- *Hotel_Ctx0 :*

```
CONTEXT
    Hotel_Ctx0
SETS
   GUEST    // les clients potentiels
   ROOM     // les chambres de l'hôtel
END
```

- *Hotel_M0 :*

**MACHINE**
    **Hotel_M0**
**SEES**
    **Hotel_Ctx0**
**VARIABLES**
    owns    // *les propriétaires des chambres*
**INVARIANTS**
    inv0_1 : owns∈ROOM⇸GUEST    // *la distribution des chambres aux clients*
**EVENTS**
    **INITIALISATION** ≙    // *Initialisation du système*
      **STATUS**
        ordinary
      **BEGIN**
        act1: owns≔∅    // *aucune réservation enregistrée*
      **END**

    **check_in** ≙    // *Gestion des réservations*
      **STATUS**
        ordinary
      **ANY**
        g
        r
      **WHERE**
        grd1: g∈GUEST    // *g est un client*
        grd2: r∈ROOM    // *r est une chambre*
        grd3 : r∉dom(owns)  // *la chambre r n'est pas réservée*
      **THEN**
        act1: owns(r)≔g    // *le client g devient le propriétaire de la chambre r*
      **END**

    **check_out** ≙    // *Teminaison de réservations*

```
        STATUS
          ordinary
        ANY
          g
          r
        WHERE
          grd1 : r↦g∈owns    //  g est le propriétaire de r
        THEN
          act1: owns≔owns\{r↦g}    //  la chambre r devient non réservée
        END

END
```

### 3.4.2  Premier raffinement : Introduction des clés électroniques et des cartes magnétiques

Il s'agit d'un modèle *moins abstrait* qui raffine le modèle initial sans le contredire. Ce modèle est plus précis en termes de détails ajoutés. En effet, on a introduit la notion de clés électroniques et de cartes magnétiques. Une telle notion favorise un accès sécurisé aux chambres de l'hôtel.

Dans ce niveau d'abstraction, nous nous intéressons aux contraintes **ENV-2** et **FUN-4** du texte référentiel :

| *L'accès* aux chambres est contrôlé par des *cartes magnétiques* | FUN-4 |

| Une carte magnétique contient *deux clés électroniques distinctes* : une *première* clé k1 modélisant le client précédent et une *deuxième* clé k2 modélisant le client actuel de la chambre | ENV-2 |

#### *3.4.2.1  Les paramètres*

Les paramètres « concrets » à ce niveau sont décrits dans un contexte nommé **Hotel_Ctx1** comme suit :

- les ensembles :

- **KEY** modélise l'ensemble des **clés potentielles**,

- Les constantes :

  - **CARD** : modélisant la notion de **carte magnétique**
  - **first** : la première clé d'une carte modélisant le **propriétaire précédent**
  - **second** : la deuxième clé d'une carte modélisant le **propriétaire actuel**
  - **f** : pour **initialiser** les clés des chambres

Les propriétés formelles relatives à ces constantes sont exprimées par les axiomes suivants :

- *axm1_1 : CARD⊆((KEY×KEY)\id)* : pour dire qu'une carte est constituée de **deux clés électroniques** non identiques

- *axm1_4 : ∀c·(c∈CARD⇒first(c)≠second(c))* : les deux clés d'une même cartes sont **différentes (distinctes)**

- *axm1_2 : first∈CARD→KEY* : rend la clé k1 modélisant le **propriétaire précédent**

- *axm1_3 : second∈CARD→KEY* : rend la clé k2 modélisant le **propriétaire actuel**

- *axm1_5 : f∈ROOM↣KEY* : f va distribuer **initialement** des clés à toutes les chambres. La fonction **f** est injective : deux chambres différentes ne peuvent pas avoir la même clé

- *axm1_6 : ran(first)=dom(CARD)* : la clé première d'une carte appartient à cette carte

- *axm1_7: ran(second)=ran(CARD)* : la clé seconde d'une carte appartient à cette carte

Ainsi, les nouveaux paramètres (concrets) introduits à ce niveau sont représentés comme suit :

**SETS**

KEY   // *les clés potentielles*

**CONSTANTS**

CARD   // *ensemble des cartes potentielles*

first   // *propriétaire précédent (clé k1)*

second   // *propriétaire actuel (clé k2)*

f   // *affectation initiale des clés pour les chambres*

**AXIOMS**

axm1_1: CARD⊆((KEY×KEY)\id)   // *une carte est constituée de deux clés*

axm1_2: first∈CARD→KEY   // *rend la première clé d'une carte*

axm1_3: second∈CARD→KEY   // *rend la deuxième clé d'une carte*

axm1_4: ∀c·(c∈CARD⇒first(c)≠second(c))   // *clés distinctes d'une même carte*

axm1_5: f∈ROOM↣KEY   // *à chaque chambre on associe une clé*

axm1_6: ran(first)=dom(CARD)

axm1_7: ran(second)=ran(CARD)

### 3.4.2.2 Etat du système

L'état concret de ce modèle est décrit par :

- Quatre *variables* :

    - **owns** : héritée du niveau abstrait

    - **cards** : **distribution** des cartes aux clients

    - **issued** : modélisant l'ensemble des **clés déjà utilisées** afin d'éviter la réutilisation des clés consommées

    - **currk** : associe à chaque chambre sa **clé courante**

- Trois propriétés invariantes notées ***inv1_1***, ***inv1_2*** et ***inv1_3*** destinées toutes au typage des variables *concrètes*.

| **VARIABLES** | **INVARIANTS** |
|---|---|
| owns | inv1_1 : cards∈CARD⇸GUEST |
| cards | inv1_2 : issued⊆KEY |
| issued | inv1_3 : currk∈ROOM↣KEY |
| currk | |

Pour l'état initial de notre système à ce niveau, il est représenté dans l'événement ***INITIALISATION*** sous forme de quatre substitutions (actions) :

**INITIALISATION** ≙    // *Initialisation du système*
  **STATUS**
    **ordinary**
  **BEGIN**
    act1: owns≔∅    // *aucune réservation enregistrée*
    act2: cards≔∅   // *cartes non encore distribuées aux clients*
    act3: issued≔ran(f) // *les clés déjà utilisées sont celles initialement affectées aux*
                              // *chambres*
    act4: currk≔f // *clés courantes pour les chambres= affectation f initiale des clés aux*
                              // *chambres*
  **END**

Les deux obligations de preuve relatives à cet événement ont été déchargées automatiquement par le prouveur de la plateforme **RODIN**.

### 3.4.2.3 Les transitions

A ce niveau, on n'introduit pas de nouvelles transitions.

Ainsi, ce modèle comporte deux *événements* (autre que l'événement INITIALISATION) *check_in1* et *check_out1* qui affinent respectivement *check_in* et *check_out*.

- *L'événement check_in1 :*

  Cet événement raffine *check_in* sans le contredire et en renforçant sa garde par un nouveau paramètre *c* modélisant une carte. Par conséquent, il admet trois paramètres *g*, *r* et *c*. Admettant que ces derniers respectent les huit gardes de l'événement –expliquées ultérieurement-, quatre substitutions vont avoir lieu modifiant l'état de ce modèle.

Voici la spécification d'un tel événement :

```
check_in1  ≙    //  Gestion des réservations
   STATUS
     ordinary
   REFINES
     check_in
   ANY
     g
     r
     c
   WHERE
     grd1: g∈GUEST    //  g est un client
     grd2: r∈ROOM    //  r est une chambre
     grd3 : r∉dom(owns)    //  la chambre r n'est pas réservée
     grd4 : c∈CARD    //  c est une carte
     grd5 : first(c)=currk(r)    //  la clé k1 correspond à la clé courante de r
     grd6 : second(c) ∉issued    //  éviter l'accès multiple à la même chambre r
     grd7 : second(c) ∉ran(currk)    //  éviter l'accès du client g à multiples chambres
     grd8 : c∉dom(cards)    //  c non associée à aucun client
   THEN
     act1: owns(r)≔g    //  le client g devient le propriétaire de la chambre r
     act2: issued≔issued∪{second(c)}    //  la clé seconde de c est marquée utilisée
     act3: cards(c)≔ g    //  la carte c est servie (délivrée) au client g
     act4: currk(r)≔second(c)    //  la clé courante de r devient la clé seconde de c
   END
```

- *L'événement check_out1 :*

Il raffine *check_out* du modèle initial. En fait, il est augmenté par un seul paramètre c, de plus ses gardes sont renforcées et de même pour ses substitutions (actions). Par conséquent, cet événement admet trois paramètres **g**, **r** et **c**, deux gardes **grd1** et **grd2** et deux actions **act1** et **act2**.

```
check_out1 ≜    // Terminaison de réservations
   STATUS
     ordinary
   REFINES
     check_out
   ANY
     g
     r
     c
   WHERE
     grd1 : r↦g∈owns    // g est le propriétaire de r
     grd2 : c↦g∈cards   // g doit être le propriétaire de c
   THEN
     act1: owns≔owns\{r↦g}    // réservation terminée
     act2 : cards≔cards\{c↦g} // carte retirée du client
END
```

### 3.4.2.4 Premier modèle raffiné

A ce niveau d'abstraction, le modèle formant l'application SCEH est composé de deux composants Event-B : un contexte **Hotel_Ctx1** qui étend le contexte *Hotel_Ctx0* et une machine **Hotel_M1** qui raffine la machine *Hotel_M0* et qui voit *explicitement* **Hotel_Ctx1** et *implicitement* le contexte **Hotel_Ctx0**.

- *Hotel_Ctx1 :*

**CONTEXT**
    Hotel_Ctx1
**EXTENDS**

| **Hotel_Ctx0** |
| --- |

**SETS**

   KEY   // *les clés potentielles*

**CONSTANTS**

   CARD   // *ensemble des cartes potentielles*

   first   // *propriétaire précédent (clé k1)*

   second   // *propriétaire actuel (clé k2)*

   f   // *affectation initiale des clés pour les chambres*

**AXIOMS**

   axm1_1: CARD⊆((KEY×KEY)\id)   // *une carte est constituée de deux clés*

   axm1_2: first∈CARD→KEY   // *rend la première clé d'une carte*

   axm1_3: second∈CARD→KEY   // *rend la deuxième clé d'une carte*

   axm1_4: ∀c·(c∈CARD⇒first(c)≠second(c))   // *clés distinctes d'une même carte*

   axm1_5: f∈ROOM⤔KEY   // *à chaque chambre on associe une clé*

   axm1_6: ran(first)=dom(CARD)

   axm1_7: ran(second)=ran(CARD)

**END**

- *Hotel_M1 :*

| **MACHINE** |
| --- |
|    **Hotel_M1** |
| **REFINES** |
|    **Hotel_M0** |
| **SEES** |
|    **Hotel_Ctx1** |
| **VARIABLES** |
|    owns |
|    cards |
|    issued |
|    currk |
| **INVARIANTS** |

    inv1_1 : cards∈CARD⇸GUEST

  inv1_2 : issued⊆KEY

  inv1_3 : currk∈ROOM⇸KEY

**EVENTS**

  **INITIALISATION** ≙   // *Initialisation du système*

    **STATUS**

      **ordinary**

    **BEGIN**

      act1: owns≔∅   // *aucune réservation enregistrée*

      act2: cards≔∅   // *cartes non encore distribuées aux clients*

      act3: issued≔ran(f)  // *les clés déjà utilisées sont celles initialement affectées*

          // *aux chambres*

      act4: currk≔f   // *clés courantes pour les chambres= affectation f initiale*

          //*des clés aux chambres*

    **END**

  **check_in1** ≙   // *Gestion des réservations*

    **STATUS**

      **ordinary**

    **REFINES**

      check_in

    **ANY**

      g

      r

      c

    **WHERE**

      grd1: g∈GUEST   // *g est un client*

      grd2: r∈ROOM   // *r est une chambre*

      grd3 : r∉dom(owns)   // *la chambre r n'est pas réservée*

      grd4 : c∈CARD   // *c est une carte*

      grd5 : first(c)=currk(r)

      grd6 : second(c)∉issued   // *éviter l'accès multiple à la même chambre r*

      grd7 : second(c)∉ran(currk)   // *éviter l'accès du client g à multiples chambres*

grd8 : c∉dom(cards)   // *c non associée à aucun client*

**THEN**

  act1: owns(r) ≔ g   // *le client g devient le propriétaire de la chambre r*

  act2: issued≔issued∪{second(c)}   // *la clé seconde de c devient utilisée*

  act3: cards(c)≔g   // *la carte c est servie au client g*

  act4: currk(r) ≔ second(c)   // *la clé courante de r devient la clé seconde de c*

**END**

**check_out1** ≙   // *Terminaison de réservations*

  **STATUS**

  **ordinary**

  **REFINES**

  check_out

  **ANY**

  g

  r

  c

  **WHERE**

  grd1 : r↦g∈owns   // *g est le propriétaire de r*

  grd2 : c↦g∈cards   // *g doit être le propriétaire de c*

  **THEN**

  act1: owns≔owns\{r↦g}   // *réservation terminée*

  act2 : cards≔cards\{c↦g}   // *carte retirée du client*

  **END**

**END**

Les 10 obligations de preuve liées à cette machine ont été déchargées automatiquement par le prouveur de la plateforme **RODIN**. Ceci prouve la correction de cette étape de raffinement (voir **figure 3.5**).

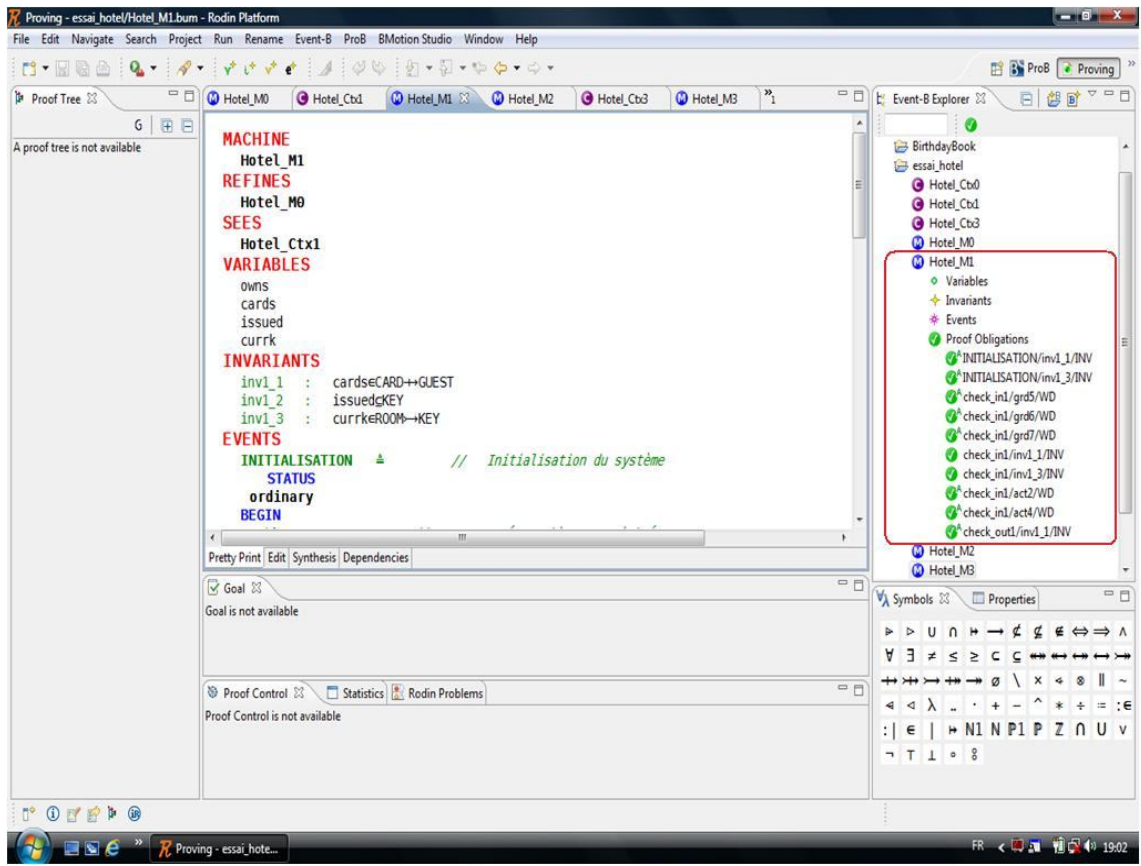

**Figure 3.5 :** *Machine Hotel_M1 avec ses obligations de preuve déchargées automatiquement*

### 3.4.3 Deuxième raffinement : Introduction des serrures électroniques

Dans cette étape de raffinement, on va introduire une nouvelle notion : les *serrures électroniques* décrites par les contraintes **ENV-1** et **FUN-3**

| | |
|---|---|
| Chaque porte est équipée d'une serrure électronique mémorisant une clé électronique et dotée d'un lecteur de cartes magnétiques | ENV-1 |
| La première visite d'un client à sa chambre est traduite par une mise à jour de la clé enregistrée dans la serrure | FUN-3 |

#### 3.4.3.1 Les paramètres

On garde les mêmes paramètres définis dans le contexte *Hotel_Ctx1*.

**SETS**

GUEST   // *les clients potentiels*

ROOM   // *les chambres de l'hôtel*

KEY   // *les clés potentielles*

**CONSTANTS**

CARD   // *ensemble des cartes potentielles*

first   // *propriétaire précédent (clé k1)*

second   // *propriétaire actuel (clé k2)*

f   // *affectation initiale des clés pour les chambres*

**AXIOMS**

axm1_1: CARD⊆((KEY×KEY)\id)   // *une carte est constituée de deux clés*

axm1_2: first∈CARD→KEY   // *rend la première clé d'une carte*

axm1_3: second∈CARD→KEY   // *rend la deuxième clé d'une carte*

axm1_4: ∀c·(c∈CARD⇒first(c)≠second(c))   // *clés distinctes d'une même carte*

axm1_5: f∈ROOM⇸KEY   // *à chaque chambre on associe une clé*

axm1_6: ran(first)=dom(CARD)

axm1_7: ran(second)=ran(CARD)

### 3.4.3.2 Etat du système

L'état concret de ce système est représenté par l'ajout d'une variable *roomk* qui indique la clé mémorisée par la serrure électronique. Cette variable est typée via un invariant noté *inv2_1*: à chaque chambre est associée une clé électronique.

**VARIABLES**

owns

cards

issued

currk

roomk

**INVARIANTS**

inv2_1 : roomk∈ROOM⇸KEY

Ainsi, l'état initial de notre système à ce niveau est décrit dans l'événement ***INITIALISATION*** provenant du modèle abstrait augmenté par la substitution ***act5 : roomk≔f***

### 3.4.3.3 Les transitions

Ce modèle est caractérisé par l'ajout de trois nouveaux événements appelés ***enter_room_change***, ***enter_room_normal*** et ***leave_room***. De tels événements modélisent respectivement les transitions : entrer dans une chambre avec changement de clé, entrer dans une chambre sans changement de clé et quitter une chambre.

On note que les versions concrètes des événements *check_in1* et *check_out1* nommées respectivement ***check_in2*** et ***check_out2*** gardent les mêmes caractéristiques que leurs abstractions avec l'ajout d'une garde ***grd9 : roomk(r)= currk(r)*** à ***check_in2*** modélisant le fait que la clé enregistrée par la serrure doit être la clé courante de la chambre (conformité des clés).

- ***L'événement enter_room_change :***

    Il s'agit d'une modélisation de l'action d'entrée d'un client dans une chambre r avec une carte c <u>pour la première fois</u>. Cela est exprimé à l'aide d'une action *act1* après vérification des quatre gardes respectives :

    - *grd1 : c∈ran(cards)* : la carte c doit être à la possession d'un client,

    - *grd2 : roomk(r)=first(c)* : la clé de la serrure est celle du propriétaire précédent,

    - *grd3 : second(c)∉ran(roomk)* : la clé seconde ne doit pas être enregistrée dans une autre serrure : éviter l'accès à multiples chambres,

    - *grd4 : r∈dom(owns)* : la chambre r doit être réservée.

L'action *act1* veut dire que la clé de la serrure est changée après une première visite. Ceci correspond exactement à la contrainte **FUN-3** de la partie référentielle du cahier de charges. Le client peut, désormais, entrer à sa chambre avec la clé seconde de sa carte c'est-à-dire sans changement de clé.

```
enter_room_change ≙    //   Client entrant dans sa chambre avec changement de clé
  STATUS
    ordinary
  ANY
    r
    c
  WHERE
    grd1: c∈dom(cards)
    grd2: roomk(r)=first(c)
    grd3 : second(c)∉ran(roomk)
    grd4 : r∈dom(owns)
  THEN
    act1: roomk(r):=second(c)
  END
```

- *L'événement enter_room_normal :*

Admettant que le client est déjà entré à la chambre r une première fois, cet événement modélise l'accès à r durant la période de réservation. Cela est traduit par une action *skip* qui ne fait rien. Un tel événement admet deux paramètres r et c qui doivent respecter trois gardes définies comme suit :

- *grd1 : c∈ran(cards)* : la carte c doit être retenue par un client,

- *grd2 : r∈dom(owns)* : la chambre r doit être réservée,

- *grd3 : roomk(r)=second(c)* : la clé seconde de c est déjà dans la serrure : il s'agit du second accès ou plus à r.

```
enter_room_normal ≙    //   Client entrant dans sa chambre sans changement de clé
  STATUS
    ordinary
  ANY
    r
    c
  WHERE
```

>     grd1: c∈dom(cards)
>
>     grd2: r∈dom(owns)
>
>     grd3 : roomk(r)≔second(c)
>
>   **THEN**
>
>     skip
>
>   **END**

- *L'événement leave_room:*

    Il modélise l'action de quitter une chambre, et il a les mêmes caractéristiques que l'événement *enter_room_normal*.

### 3.4.3.4 Deuxième modèle raffiné

Ci-dessous la version concrète *Hotel_M2* qui affine la machine abstraite *Hotel_M1* :

> **MACHINE**
>     **Hotel_M2**
> **REFINES**
>     **Hotel_M1**
> **SEES**
>     **Hotel_Ctx1**
> **VARIABLES**
>     owns
>     cards
>     issued
>     currk
>     roomk
> **INVARIANTS**
>     inv2_1  : roomk∈ROOM⇸KEY
> **EVENTS**
>     **INITIALISATION** ≙   //  *Initialisation du système*

**STATUS**

  **ordinary**

**BEGIN**

  act1: owns≔∅   // *aucune réservation enregistrée*

  act2: cards≔∅   // *cartes non encore distribuées aux clients*

  act3: issued≔ran(f)  // *les clés déjà utilisées sont celles initialement affectées*
                      // *aux chambres*

  act4: currk≔f    // *clés courantes pour les chambres= affectation f initiale*
                      //*des clés aux chambres*

  act5: roomk≔f

**END**

**check_in2** ≜  // *Gestion des réservations*

  **STATUS**

    **ordinary**

  **REFINES**

    check_in1

  **ANY**

    g

    r

    c

  **WHERE**

    grd1: g∈GUEST   // *g est un client*

    grd2: r∈ROOM   // *r est une chambre*

    grd3 : r∉dom(owns)   // *la chambre r n'est pas réservée*

    grd4 : c∈CARD   // *c est une carte*

    grd5 : first(c)=currk(r)

    grd6 : second(c)∉issued   // *éviter l'accès multiple à la même chambre r*

    grd7 : second(c)∉ran(currk)   // *éviter l'accès du client g à multiples chambres*

    grd8 : c∉dom(cards)   // *c non associée à aucun client*

    grd9 : roomk(r)= currk(r)

  **THEN**

    act1: owns(r) ≔g   // *le client g devient le propriétaire de la chambre r*

    act2: issued≔issued∪{second(c)}   // *la clé seconde de c devient utilisée*

act3: cards(c)≔g    // *la carte c est servie au client g*

    act4: currk(r) ≔second(c)    // *la clé courante de r devient la clé seconde de c*

  **END**

**check_out2** ≙    // *Terminaison de réservations*

  **STATUS**

   ordinary

  **REFINES**

   check_out1

  **ANY**

   g

   r

   c

  **WHERE**

   grd1 : r↦g∈owns    // *g est le propriétaire de r*

   grd2 : c↦g∈cards    // *g doit être le propriétaire de c*

  **THEN**

   act1: owns≔owns\{r↦g}    // *réservation terminée*

   act2 : cards≔cards\{c↦g}    // *carte retirée du client*

  **END**

**enter_room_change** ≙    // *Client entrant dans sa chambre avec changement de clé*

  **STATUS**

   ordinary

  **ANY**

   r

   c

  **WHERE**

   grd1: c∈dom(cards)

   grd2: roomk(r)=first(c)

   grd3 : second(c)∉ran(roomk)

   grd4 : r∈dom(owns)

  **THEN**

   act1: roomk(r)≔second(c)

  **END**

**enter_room_normal** ≜   *//  Client entrant dans sa chambre sans changement de clé*
  **STATUS**
   ordinary
  **ANY**
   r
   c
  **WHERE**
   grd1: c∈dom(cards)
   grd2: r∈dom(owns)
   grd3 : roomk(r)≔second(c)
  **THEN**
   skip
  **END**

**leave_room** ≜   *//  Client quittant sa chambre*
  **STATUS**
   ordinary
  **ANY**
   r
   c
  **WHERE**
   grd1: r∈dom(owns)
   grd2: c∈dom(cards)
   grd3 : roomk(r)≔second(c)
  **THEN**
   skip
  **END**

**END**

Les 8 obligations de preuve liées à cette machine ont été déchargées automatiquement par le prouveur de la plateforme **RODIN**. Ce qui prouve la correction de cette étape de raffinement.

### 3.4.4 Troisième raffinement : Introduction des employés de l'hôtel

C'est le modèle le plus concret de notre modélisation de l'application SCEH. Il décrit la spécification formelle en Event-B de l'application SCEH.

#### 3.4.4.1 Les paramètres

En se basant sur la contrainte **FUN-2**, on peut extraire un nouveau paramètre **ADMINISTRATOR** pour noter l'ensemble des employés de l'hôtel. A chacun de ces employés, on a affecté un ensemble de chambres. Ceci est exprimé à l'aide d'une fonction constante **owns_adm** ayant ROOM et ADMINISTRATOR comme ensemble de départ et d'arrivée respectivement (voir **axm3_1**).

Ces paramètres sont rassemblés dans le contexte *Hotel_Ctx3* représenté ci-dessous :

```
CONTEXT
    Hotel_Ctx3
EXTENDS
    Hotel_Ctx1
SETS
    ADMINISTRATOR   // les administrateurs potentiels
CONSTANTS
    owns_adm    //  affectation des administrateurs aux chambres
AXIOMS
    axm3_1: owns_adm ∈ ROOM→ADMINISTRATOR
END
```

#### 3.4.4.2 Etat du système

L'état concret de ce système est représenté par l'ajout d'une variable **cards_adm** qui est une *fonction partielle au sens mathématique* distribuant les cartes aux administrateurs : ceci est exprimé au niveau d'un invariant **inv3_1**.

Ainsi, l'état du système à ce niveau est représenté comme suit :

```
VARIABLES
    owns
    cards
    issued
    currk
    roomk
    cards_adm
INVARIANTS
    inv3_1  :  cards_adm ∈ CARD ⇸ ADMINISTRATOR
```

Pour l'état initial de notre système, il est caractérisé dans l'événement *INITIALISATION* par l'ajout de la substitution ***act6 : cards_adm ≔ ∅*** : cartes non distribuées aux personnels.

### 3.4.4.3  Les transitions

A ce niveau, le modèle est enrichi par deux nouvelles transitions relatives aux employés étiquetées ***enter_room_normal_adm*** et ***leave_room_adm*** et désignant respectivement l'action d'entrée à une chambre et de sortie d'une chambre par un personnel de l'hôtel. Ces deux transitions ont presque les mêmes signatures que *enter_room_normal3* et *leave_room3*. De plus, les transitions du modèle abstrait sont conservées dans ce modèle tout en intégrant quelques aspects liés à la notion administrateur tels que l'ajout du paramètre de type ADMINISTRATOR et des gardes liées à ce paramètre dans les événements ***check_in3*** et ***check_out3*** versions concrètes des événements *check_in2* et *check_out2*. Les autres événements restent inchangeables à savoir ***enter_room_change3***, ***enter_room_normal3*** et ***leave_room3***.

### 3.4.4.4  Troisième modèle raffiné

Ci-dessous, la spécification Event-B de la dernière machine appelée ***Hotel_M3*** qui raffine la machine abstraite *Hotel_M2* et voit le dernier context ***Hotel_Ctx3*** ainsi que ses ancêtres.

**MACHINE**
    **Hotel_M3**
**REFINES**
    **Hotel_M2**
**SEES**
    **Hotel_Ctx3**
**VARIABLES**
    owns
    cards
    issued
    currk
    roomk
    cards_adm
**INVARIANTS**
    inv3_1 : cards_adm ∈ CARD ⇸ ADMINISTRATOR
**EVENTS**
    **INITIALISATION** ≜    // *Initialisation du système*
      **STATUS**
        **ordinary**
      **BEGIN**
        act1: owns≔∅    // *aucune réservation enregistrée*
        act2: cards≔∅    // *cartes non encore distribuées aux clients*
        act3: issued≔ran(f)  // *les clés déjà utilisées sont celles initialement affectées aux*
                     // *chambres*
        act4: currk≔f      // *clés courantes pour les chambres= affectation f initiale*
                      // *des clés aux chambres*
        act5: roomk≔f
        act6: cards_adm≔∅    // *cartes non encore distribuées aux personnels*
      **END**

    **check_in3** ≜    // *Gestion des réservations*
      **STATUS**
        **ordinary**

**REFINES**

  check_in2

**ANY**

  g

  r

  c

  a

**WHERE**

  grd1: g∈GUEST   // *g est un client*

  grd2: r∈ROOM   // *r est une chambre*

  grd3 : r∉dom(owns)   // *la chambre r n'est pas réservée*

  grd4 : c∈CARD   // *c est une carte*

  grd5 : first(c)=currk(r)

  grd6 : second(c)∉issued   // *éviter l'accès multiple à la même chambre r*

  grd7 : second(c)∉ran(currk)   // *éviter l'accès du client g à multiples chambres*

  grd8 : c∉dom(cards)   // *c non associée à aucun client*

  grd9 : roomk(r) = currk(r)

  grd10 : a∈ADMINISTRATOR

  grd11 : owns_adm(r)=a

  grd12 : c∉dom(cards_adm)

**THEN**

  act1: owns(r) ≔g   // *le client g devient le propriétaire de la chambre r*

  act2: issued≔issued∪{second(c)}   // *la clé seconde de c devient utilisée*

  act3: cards(c)≔ g   // *la carte c est servie au client g*

  act4: currk(r) ≔second(c)   // *la clé courante de r devient la clé seconde de c*

  act5: cards_adm(c)≔a

**END**

**check_out3** ≙  // *Terminaison de réservation*

  **STATUS**

    ordinary

  **REFINES**

    check_out2

  **ANY**

  g

  r

  c

**WHERE**

 grd1 : r↦g∈owns *// g est le propriétaire de r*

 grd2 : c↦g∈cards *// g doit être le propriétaire de c*

 grd3 : c∈dom(cards_adm) *// c est à la propriété d'un administrateur*

**THEN**

 act1: owns≔owns\{r↦g} *// réservation terminée*

 act2 : cards≔cards\{c↦g} *// carte retirée du client*

 act3 : cards_adm≔{c}⩤cards_adm *// carte retirée de l'administrateur approprié*

**END**

**enter_room_change3** ≙ *// Client entrant dans sa chambre avec changement de clé*

 **STATUS**

  **ordinary**

 **REFINES**

  enter_room_change

 **ANY**

  r

  c

 **WHERE**

  grd1: c∈dom(cards)

  grd2: roomk(r)=first(c)

  grd3 : second(c)∉ran(roomk)

  grd4 : r∈dom(owns)

 **THEN**

  act1: roomk(r)≔second(c)

 **END**

**enter_room_normal3** ≙ *// Client entrant dans sa chambre sans changement de clé*

 **STATUS**

  **ordinary**

 **REFINES**

  enter_room_normal

    **ANY**
      r
      c
    **WHERE**
      grd1: c∈dom(cards)
      grd2: r∈dom(owns)
      grd3 : roomk(r)≔second(c)
    **THEN**
      skip
    **END**

**leave_room3** ≜   //   *Client quittant sa chambre*
    **STATUS**
      **ordinary**
    **REFINES**
      leave_room
    **ANY**
      r
      c
    **WHERE**
      grd1: r∈dom(owns)
      grd2: c∈dom(cards)
      grd3 : roomk(r)≔second(c)
    **THEN**
      skip
    **END**

**enter_room_normal_adm** ≜   //   *Personnel entrant dans une chambre*
    **STATUS**
      **ordinary**
    **ANY**
      r
      c
    **WHERE**
      grd1: c∈dom(cards_adm)
      grd2: r∈dom(owns)

```
            grd3 : roomk(r)≔second(c)
        THEN
           skip
        END
     leave_room_adm ≙    //   Personnel quittant une chambre
        STATUS
           ordinary
        ANY
           r
           c
        WHERE
           grd1 : r∈dom(owns)
           grd2 : c∈dom(cards_adm)
            grd3 : roomk(r)≔second(c)
        THEN
           skip
        END

END
```

Les 6 obligations de preuve relatives à cette machine ont été déchargées automatiquement par le prouveur de la plateforme **RODIN**.

Ce dernier modèle traduit la spécification formelle en Event-B de l'application SCEH. La construction de cette spécification formelle est obtenue suite à l'application de la stratégie de raffinement horizontal proposée en **3.3**. Toutes les preuves liées aux propriétés de sûreté (préservation de l'invariant et correction du raffinement) ont été déchargées (ou démontrées). Ainsi, on peut dire que cette spécification formelle de l'application SCEH est correcte par construction.

La **table 3.1** récapitule la prise en compte de toutes les contraintes issues du cahier des charges structuré de l'application SCEH.

| Modèle | FUN | ENV |
|---|---|---|
| Initial (*Hotel_Ctx0*, *Hotel_M0*) | FUN-1 | _ |
| Premier raffinement (*Hotel_Ctx1*, *Hotel_M1*) | FUN-4 | ENV-2 |
| Deuxième raffinement (*Hotel_M2*) | FUN-3 | ENV-1 |
| Troisième raffinement (*Hotel_Ctx3*, *Hotel_M3*) | FUN-2 | _ |

**Table 3.1 :** *Table de synthèse de la stratégie de raffinement proposée*

## 3.5   Animation de la spécification formelle

Nous avons validé les modèles Event-B construits dans ce travail en menant une activité d'animation supportée par l'outil ProB. Dans la suite, nous nous limitons à trois scénarios d'animation de la spécification formelle de l'application SCEH.

- **Scénario 1 : Autant de check_in que de check_out**

    En partant d'un état initial de la machine abstraite Hotel_M0, où aucune réservation n'est enregistrée, et moyennant un nombre *n* de réservations (n>0) et un nombre *m* de terminaisons de réservation (m>0) avec *n=m*, on doit regagner l'état initial (aucune réservation).

    L'enchaînement des événements décrivant ce scénario est :

    | |
    |---|
    | *check_in(GUEST1, ROOM1)* → *check_in(GUEST2, ROOM2)* → *check_out(GUEST2, ROOM2)* → *check_out(GUEST1, ROOM1)*     *(scénario1)* |

    Nous avons vérifié les comportements dynamiques de ce scénario avec ProB. Sachant que celui-ci choisit les valeurs des ensembles porteurs, des constantes et des paramètres des événements à franchir. En outre, nous avons vérifié le non-blocage du modèle **Hotel_M0** avec ProB. De plus, ProB n'a pas trouvé d'état violant **inv0_1** de la machine **Hotel_M0**.

- **Scénario 2 : Tentative de réservation d'une chambre occupée**

En principe, le scénario donné ci-dessous ne devrait pas être supporté par le modèle **Hotel_M0**. En effet, la chambre ROOM1 est affectée à deux clients différents (GUEST1 et GUEST3). Mais lors de l'exécution de ce scénario, nous avons remarqué que l'invariant **inv0_1** de la machine Hotel_M0 est toujours vérifié ! Après avoir analysé la machine Hotel_M0, nous avons pu localiser l'erreur. La garde grd3 (**r↦g∉owns**) de l'événement check_in est coupable. Elle est corrigée comme suit : **grd3 : r∉dom(owns)**. Après correction de cette erreur, nous avons pu prouver la nouvelle machine Hotel_M0. Et nous avons vérifié que le scénario 2 n'est pas supporté par la machine Hotel_M0 corrigée (voir **Figure 3.6**).

*check_in(GUEST1, ROOM1)→ check_in(GUEST2, ROOM2)→ check_in(GUEST3, ROOM1) → check_out(GUEST2, ROOM2) → check_out(GUEST3, ROOM1)* *(scénario2)*

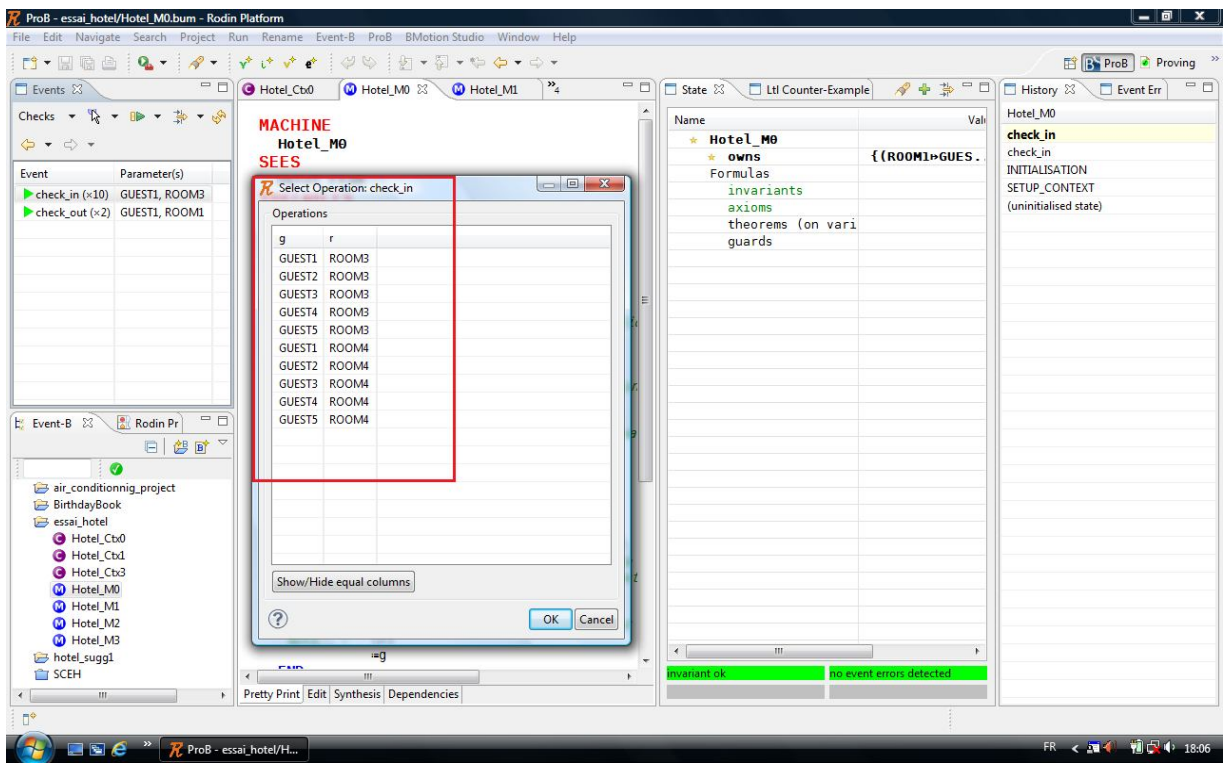

**Figure 3.6 :** *Animation du premier modèle (machine abstraite Hotel_M0) par l'outil ProB*

- **Scénario 3 : Le non-blocage du modèle Hotel_M1**

En Event-B, le non-blocage en tant que propriété de vivacité peut être prouvé par la démonstration d'un théorème dont la formule est la disjonction des gardes des événements du modèle concerné. Le théorème d'interblocage de la machine Hotel_M1 est :

$\exists g,r,c \cdot (g \in GUEST \land r \in ROOM \land r \notin dom(owns) \land c \in CARD \land first(c) = currk(r)$
$\land second(c) \notin issued \land second(c) \notin ran(currk) \land c \notin dom(cards)) \lor \exists g1,r1,c1 \cdot$
$(r1 \mapsto g1 \in owns \land c \mapsto g \in cards))$

Nous n'avons pas pu décharger ce théorème. Pour faire face à ce problème, nous avons utilisé le model-checker ProB. Son lancement sur le modèle Hotel_M1 donne :

```
Deadlock found!
  Deadlock found.
ProB has detected a state where all guards are false.
The following is the trace that led to the deadlock:
'SETUP_CONSTANTS({(KEY1|->KEY1)},{(ROOM1|->KEY1),(ROOM2|->KEY2)},{((KEY1|-
>KEY1)|->KEY1)},{((KEY1|->KEY1)|->KEY2)})'
'INITIALISATION({},{(ROOM1|->KEY1),(ROOM2|->KEY2)},{KEY1,KEY2},{})'

=================================================
 Coverage statistics:
 Total Number of Nodes:8
 Total Number of Transitions:7
 Node Statistics:
 'invariant_violated:0'
 'explored_but_not_all_transitions_computed:1'
 'live:4'
 'deadlocked:1'
 'open:3'
 'invariant_not_checked:3'
 'total:8'
 Operations Statistics:
 '$setup_constants:4'
 '$initialise_machine:3'
 Uncovered Operations:
 check_in1
 check_out1
```

En analysant les résultats fournis par ProB, nous constatons les faits non attendus suivants :

- Une carte peut contenir deux clés identiques (CARD={(KEY1|->KEY1)}),
- Les deux fonctions *first* et *second* sont assimilées à deux fonctions indépendantes.

Ces anomalies peuvent être corrigées en ajoutant des axiomes **explicitant** les propriétés des paramètres de l'application SCEH tels que :

**axm1_1 : CARD⊆((KEY×KEY)\id)**

**axm1_6 : ran(first)=dom(CARD)**

**axm1_7 : ran(second)=ran(CARD)**

## 3.6  Conclusion

Dans ce chapitre, nous avons construit pas-à-pas une spécification formelle en Event-B de l'application *SCEH*. Pour y parvenir, nous avons élaboré une stratégie de raffinement horizontal adaptée à l'application SCEH. Une telle stratégie a été appliquée avec succès et a donné naissance à une machine abstraite, trois machines raffinées et trois contextes. En outre, nous avons validé la spécification formelle obtenue en utilisant avec profit le model-checker ProB.

Dans le chapitre suivant, nous allons transformer la spécification formelle en Event-B de l'application SCEH vers un modèle UML/OCL en passant par un modèle UML/EM-OCL pivot.

# Chapitre 4 : D'une spécification en Event-B de l'application SCEH vers un diagramme de classes UML/OCL en passant par UML/EM-OCL

## 4.1 Introduction

Dans ce chapitre, nous proposons une approche permettant de transformer la spécification formelle en Event-B de l'application SCEH en un diagramme de classes UML/OCL. Pour y parvenir, le modèle ultime Hotel_M3 en Event-B de l'application SCEH est aplati : une machine Event-B (sans *refines*) et un contexte (sans *extends*). Ensuite, nous proposons des règles de transformation d'Event-B vers UML/EM-OCL. Enfin, nous transformons le modèle UML/EM-OCL obtenu en son équivalent UML/OCL. Ce chapitre comporte quatre sections. La première section permet d'aplatir le modèle ultime issu de la phase de raffinement horizontal (voir chapitre 3). La deuxième section propose des règles de transformation d'Event-B vers UML/EM-OCL et applique ces règles sur la modélisation de l'application SCEH. La troisième section propose des règles de raffinement d'UML/EM-OCL vers UML/OCL et applique ces règles sur le modèle UML/EM-OCL de l'application SCEH. Enfin, la quatrième section introduit des classes dites classes intermédiaires (helpers) au modèle UML/OCL de l'application SCEH.

## 4.2 Modèle aplati en Event-B de l'application SCEH

Dans la suite, nous présentons une version aplatie du dernier modèle Event-B de l'application SCEH. Un tel modèle aplati comporte un contexte (voir **4.2.1**) et une machine (voir **4.2.2**).

### 4.2.1 Contexte aplati

```
CONTEXT
    Hotel_Ctx3
SETS
    GUEST
    ROOM
```

```
    KEY
    ADMINISTRATOR
CONSTANTS
    CARD
    first
    second
    f
    owns_adm
AXIOMS
    axm1_1: CARD⊆((KEY×KEY)\id)
    axm1_2: first∈CARD→KEY
    axm1_3: second∈CARD→KEY
    axm1_4: ∀c·(c∈CARD⇒first(c)≠second(c))
    axm1_5: f∈ROOM⤔KEY
    axm1_6: ran(first)=dom(CARD)
    axm1_7: ran(second)=ran(CARD)
    axm3_1: owns_adm∈ROOM→ADMINISTRATOR

END
```

### 4.2.2 Machine aplatie

```
MACHINE
    Hotel_M3
SEES
    Hotel_Ctx3
VARIABLES
    owns
    cards
    issued
    currk
    roomk
    cards_adm
```

**INVARIANTS**

    inv0_1 : owns∈ROOM⇸GUEST

    inv1_1 : cards∈CARD⇸GUEST

    inv1_2 : issued⊆KEY

    inv1_3 : currk∈ROOM⇸KEY

    inv2_1 : roomk∈ROOM⇸KEY

    inv3_1 : cards_adm∈ CARD⇸ADMINISTRATOR

**EVENTS**

    **INITIALISATION** ≙    *// Initialisation du système*

      **STATUS**

        **ordinary**

      **BEGIN**

        act1: owns≔∅

        act2: cards≔∅

        act3: issued≔ran(f)

        act4: currk≔f

        act5: roomk≔f

        act6: cards_adm≔∅

      **END**

    **check_in3** ≙    *// Gestion des réservations*

      **STATUS**

        **ordinary**

      **ANY**

        g

        r

        c

        a

      **WHERE**

        grd1: g∈GUEST

        grd2: r∈ROOM

        grd3 : r∉dom(owns)

        grd4 : c∈CARD

    grd5 : first(c)=currk(r)

    grd6 : second(c)∉issued

    grd7 : second(c)∉ran(currk)

    grd8 : c∉dom(cards)

    grd9 : roomk(r)= currk(r)

    grd10 : a∈ADMINISTRATOR

    grd11 : owns_adm(r)=a

    grd12 : c∉dom(cards_adm)

  **THEN**

    act1: owns(r) ≔g

    act2: issued≔issued∪{second(c)}

    act3: cards(c)≔ g

    act4: currk(r) ≔second(c)

    act5: cards_adm(c)≔a

  **END**

**check_out3** ≙   //  *Terminaison de réservation*

  **STATUS**

    **ordinary**

  **ANY**

    g

    r

    c

  **WHERE**

    grd1 : r↦g∈owns

    grd2 : c↦g∈cards

    grd3 : c∈dom(cards_adm)

  **THEN**

    act1: owns≔owns\{r↦g}

    act2 : cards≔cards\{c↦g}

    act3  : cards_adm≔{c}⩤cards_adm

  **END**

**enter_room_change3** ≙   //  *Client entrant dans sa chambre avec changement de clé*

**STATUS**
     **ordinary**
   **ANY**
     r
     c
   **WHERE**
     grd1: c∈dom(cards)
     grd2: roomk(r)=first(c)
     grd3 : second(c)∉ran(roomk)
     grd4 : r∈dom(owns)
   **THEN**
     act1: roomk(r)≔second(c)
   **END**
 **enter_room_normal3** ≜   *//  Client entrant  dans sa chambre sans changement de clé*
   **STATUS**
     **ordinary**
   **ANY**
     r
     c
   **WHERE**
     grd1: c∈dom(cards)
     grd2: r∈dom(owns)
     grd3 : roomk(r)≔second(c)
   **THEN**
     skip
   **END**
 **leave_room3** ≜   *//  Client quittant sa  chambre*
   **STATUS**
     **ordinary**
   **ANY**
     r
     c
   **WHERE**
     grd1: r∈dom(owns)

  grd2: c∈dom(cards)

  grd3 : roomk(r)≔second(c)

 **THEN**

  skip

 **END**

**enter_room_normal_adm** ≙  //  *Personnel entrant dans une chambre*

 **STATUS**

  **ordinary**

 **ANY**

  r

  c

 **WHERE**

  grd1: c∈dom(cards_adm)

  grd2: r∈dom(owns)

  grd3 : roomk(r)≔second(c)

 **THEN**

  skip

 **END**

**leave_room_adm** ≙  //  *Personnel quittant une chambre*

 **STATUS**

  **ordinary**

 **ANY**

  r

  c

 **WHERE**

  grd1 : r∈dom(owns)

  grd2 : c∈dom(cards_adm)

  grd3 : roomk(r)≔second(c)

 **THEN**

  skip

 **END**

**END**

## 4.3 Les règles de transformation systématique d'Event-B vers UML/EM-OCL

Dans cette section, nous proposons des règles de transformation systématique d'Event-B vers UML/EM-OCL. Ces règles sont au nombre de **15**. En outre, nous montrons la faisabilité de ces règles sur le modèle Event-B de l'application SCEH.

*Règle 1: Classe fondamentale*

La dernière machine aplatie Event-B est traduite en une *classe* au sens OO.

- Application : la machine Hotel_M3 est traduite en une classe **Hotel** (pour des raisons de lisibilité, on a changé le nom (Hotel au lieu de Hotel_M3)).

*Règle 2 : Types de données (Data Type)*

Les ensembles (*SETS*) appartenant au contexte aplati vu par la dernière machine sont traduits en types de données abstraits « **Abstract Data Type** » en premier temps, et en attributs de type Set(Type-de-données-correspondant) en deuxième temps. Ces attributs sont de la forme:

 Nom_type_de_données :Set(TYPE_DE_DONNEES).

On parle ainsi des *ensembles potentiels* (Data Type) et des *ensembles effectifs* (attributs). De plus, une contrainte « *not empty* » s'ajoute à l'ensemble des contraintes EM-OCL liées au diagramme traduisant la propriété invariante des ensembles en Event-B.

- Illustration : l'ensemble abstrait *GUEST* qui est vu par la machine Hotel_M3 aplatie est transformé en Data Type *GUEST* en premier temps, puis en attribut *Guest* ayant comme type *Set(GUEST)*

*Règle 3: Attributs statiques*

Une constante est traduite soit en *attribut statique (attribut de classe)* soit en *opération de consultation (query)*. Par exemple, les constantes *CARD*, *first*, *second*, *f* et *owns_adm* sont traduites toutes en attributs de classe dans la modélisation en UML/EM-OCL. Le typage de tels attributs se réalise en se référant à la **règle 5**.

*Règle 4: Attributs Objet*

Une variable est traduite soit en *attribut objet* soit en *opération de consultation (query)*. Le typage de tels attributs se réalise en se référant à la **règle 6**.

*Règle 5: Typage des attributs statiques et Invariants*

- Un axiome (clause **AXIOMS**) contenant le symbole d'appartenance ($\in$) est uniquement destiné au <u>typage des attributs de classe</u>,

- Un axiome (clause **AXIOMS**) contenant le symbole d'inclusion ($\subseteq$) est destiné au <u>typage des attributs de classe</u> en premier temps, et à leur description sous forme de <u>contrainte EM-OCL</u> de type *invariant (inv)* en deuxième temps.

- Autrement, l'axiome est traduit par une contrainte de type *invariant*.

*Règle 6: Typage des attributs objet et Invariants*

- Un invariant (clause **INVARIANTS**) contenant le symbole d'appartenance ($\in$) est uniquement destiné au <u>typage des attributs objet</u>

- Un invariant (clause **INVARIANTS**) contenant le symbole d'inclusion ($\subseteq$) est destiné au <u>typage des attributs objet</u> en premier temps, et à leur description sous forme de <u>contrainte EM-OCL</u> de type *invariant (inv)* en deuxième temps.

- Autrement, l'invariant est traduit par une contrainte de type *invariant*.

*Règle 7: Constructeur*

L'événement particulier **INITIALISATION** est traduit en un *constructeur* stéréotypé par «*constructor*» dans la classe fondamentale.

*Règle 8: Méthodes/Opérations applicables*

Un événement spécifié dans une machine en Event-B se traduit par une *opération de modification « update »* en UML/EM-OCL dont les paramètres sont ceux de l'événement situés dans la clause **ANY**. Ces paramètres sont typés en se référant aux gardes associées à l'événement. Par exemple, l'événement check_in3 est traduit en une opération de modification donnée ci-dessous dans la classe Hotel avec *g*, *r*, *c* et *a* comme paramètres.

```
    check_in3 ≙
        .
        .
      ANY
        g
        r
        c
        a
      WHERE
        grd1: g∈GUEST
        grd2: r∈ROOM
        .
        grd4 : c∈CARD
        .
        .
        grd10 :
        a∈ADMINISTRATOR

      THEN
        .
        .
      END
```

**Règle 8** ⟹ <<update>> check_in3(g: GUEST, r: ROOM, c: Pair(Key,Key), a: ADMINISTRATOR)

*Règle 9: Préconditions extraites des gardes*

Une garde peut être utilisée soit pour le *typage des paramètres* (**Table 4.2**) de la méthode correspondante à l'évènement adéquat, soit comme une *précondition* « *pre* » ayant pour objet ces paramètres.

*Règle 10: Postconditions extraites des substitutions*

Dans un évènement en Event-B, chaque substitution est convertie en une *postcondition* « *post* ».

*Règle 11: Substitution skip*

Une action qui ne fait rien (**skip**) est traduite en une *postcondition* dont l'état de l'objet demeure inchangé.

*Règle 12: Visibilité des attributs et des méthodes*

1. Le statut d'exportation des attributs est *privé* (*private,* « - »)
2. Le statut d'exportation des méthodes est *public* (« + »)

*Règle 13: Passage des gardes implicites vers des contraintes explicites*

Un calcul explicite de typage des paramètres est parfois nécessaire en tenant compte du niveau global (état de la machine et de ses contextes)

*Règle 14: Contraintes EM-OCL*

Un axiome ou invariant contenant un ou plusieurs symboles parmi ∀, ∃, ▷, ×, ~, ; , ⇒, etc, est directement traduit en une contrainte EM-OCL contenant ces symboles traduits selon la **Table 4.1**.

| Symbole Event-B | Traduction en EM-OCL |
|---|---|
| ∀, ∃ | forAll, exists |
| ◁, ▷ | restrictionDomain(), restrictionRange() |
| ⩤, ⩥ | soustractionDomain(), soustractionRange() |
| A⇒B | implies |
| A⇔B | A **implies** B **and** B **implies** A |
| ~ | inverse() |
| r1;r2 | r1->seqComposition(r2) |
| +, −, *, ÷ | +, −, *, ÷ |
| ≠, ≤, ≥ | <>, <=, >= |
| ⊂, ⊆, ∈, ⊄, ⊈, ∉ | includes / includesAll, excludes / excludesAll |
| ∅ | isEmpty() |
| ⊗, ∥ | directProduct(), ParallelProduct() |
| ∩, ∪ | union(), intersection() |
| x↦y | Pair(x, y) |
| A\B | A->excluding(B) |
| × | product() |
| ¬ | not |

**Table 4.1 :** *Correspondances entre les symboles en Event-B et leurs traductions en EM-OCL*

*Règle 15: Correspondances de typage entre Event-B et UML/EM-OCL*

Pour les invariants, les axiomes et les gardes destinés au typage, nous avons établi des correspondances entre Event-B et UML/EM-OCL regroupées dans la **table 4.2**.

Soit x une constante, variable, ou paramètre d'un évènement, et soient A et B deux ensembles ;

| Typage en Event-B | Typage en UML/EM-OCL |
|---|---|
| x∈A | x :A |
| A⊆B | A :Set(B) |
| x∈A↔B | x :BinaryRelation(A,B) |
| x∈A⇸B | x :PartialFunction(A,B) |
| x∈ A→B | x :TotalFunction(A,B) |
| x∈ A⤔B | x :PartialInjective(A,B) |
| x∈ A↣B | x :TotalInjective(A,B) |
| x∈ A⤀B | x :PartialSurjective(A,B) |
| x∈ A↠B | x :TotalSurjective(A,B) |
| x∈ A⤖B | x :TotalBijective(A,B) |

**Table 4.2 :** *Correspondances entre le typage en Event-B et le typage en UML/EM-OCL*

Pour des raisons de **traçabilité**, nous avons intérêt à conserver les étiquettes attachées aux éléments en Event-B comme étiquettes reliées aux contraintes EM-OCL. La **table 4.3** récapitule les correspondances macroscopiques entre Event-B et UML/EM-OCL.

| **Spécification en Event-B** | **Modélisation en UML/EM-OCL** |
|---|---|
| *Partie statique* ||
| Constants + Sets + Variables | Attributs de classe + attributs Objet |
| Axioms + Invariants | Propriétés invariantes « *inv* » attachées au modèle |
| *Partie dynamique* ||
| Events | Constructeur « constructor » + Opérations de modification « update » |
| Gardes + Substitutions | Contraintes pre/post stéréotypées « pre » et « post » attachées au modèle |

**Table 4.3 :** *Correspondances macroscopiques entre Event-B en UML/EM-OCL*

## 4.4 Diagramme de classes en UML/EM-OCL de l'application SCEH

### 4.4.1 Diagramme de classes

La **figure 4.1** donne le diagramme de classes UML/EM-OCL obtenu à partir du modèle Event-B de l'application SCEH en appliquant les règles proposées dans **4.3**.

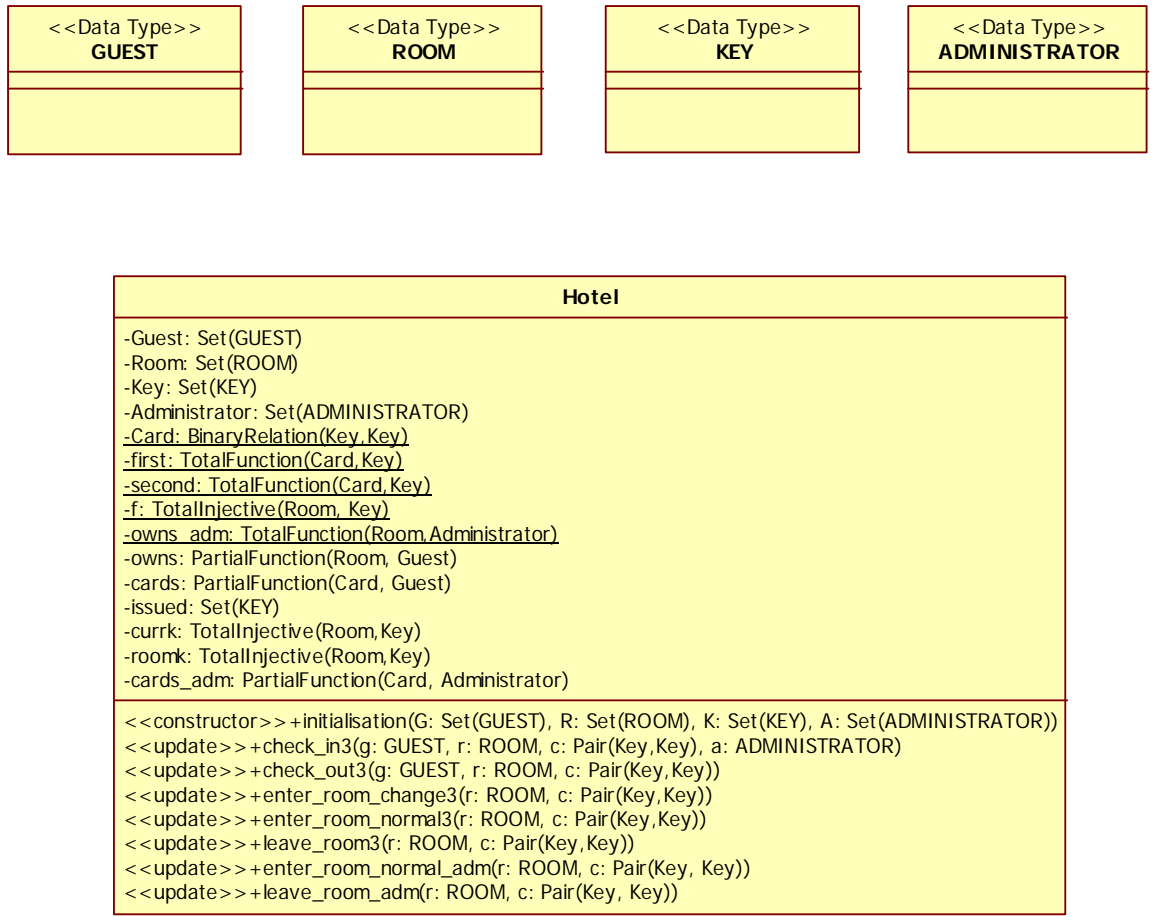

**Figure 4.1 :** *Transformation de la spécification en Event-B du SCEH vers un modèle UML/EM-OCL*

Ce diagramme est composé de quatre types de données abstraits (TDA) représentées sous forme de classes avec le stéréotype « Data Type ». Ces TDA sont notés GUEST, ROOM, KEY et ADMINISTRATOR. Ils représentent dans l'ordre les ensembles potentiels des clients, des chambres, des clés et des personnels. De plus, ce diagramme est essentiellement basé sur une classe fondamentale notée ***Hotel***. La partie statique (ou encore les attributs) d'une telle classe étant marquée *privée* « - », tandis que sa partie dynamique contenant les services offerts est à usage *public* « + ».

### 4.4.2 Contraintes attachées au modèle UML/EM-OCL

Le diagramme de classes UML/EM-OCL est insuffisant pour modéliser la spécification en Event-B du modèle SCEH : il reste des propriétés à traduire sous forme de contraintes attachées au modèle UML/EM-OCL. Dans ce qui suit, nous décrivons les contraintes EM-OCL attachées à la classe Hotel. Elles sont accompagnées de commentaires de la forme :

*--explication de la contrainte : contrainte originale venant de l'application en Event-B*.

*-- Propriétés invariantes du système*

**context** Hotel

*--propriétés des ensembles en Event-B : un ensemble abstrait est par défaut non vide*

**inv not_empty_guest** : Guest->notEmpty()

**inv not_empty_room** : Room->notEmpty()

**inv not_empty_key** : Key->notEmpty()

**inv not_empty_admin** : Administrator->notEmpty()

*-- une carte est le produit cartésien de deux clés : CARD⊆KEY×KEY*

**inv axm1_1:** (Key->product(Key)) ->includesAll(Card))

*-- les deux clés d'une carte sont différentes : ∀c·(c∈CARD⇒first(c)≠second(c))*

**inv axm1_4:** Card->forAll(c: Pair(Key, Key) | c.first()<>c.second())

*-- la clé première d'une carte appartient à cette carte : ran(first)=dom(CARD)*

**inv axm1_6:** first->range()=Card->domain()

*-- la clé seconde d'une carte appartient à cette carte : ran(second)=ran(CARD)*

**inv axm1_7:** second->range()=Card->range()

*-- l'ensemble des clés utilisées est inclus dans l'ensemble des clés potentielles : issued⊆KEY*

**inv inv1_2:** Key->includesAll(issued)

*-- Initialisation du système*

**context** Hotel:: initialisation(G: Set(GUEST), R: Set(ROOM), K: Set(KEY), A: Set(ADMINISTRATOR))

**pre:** G->notEmpty()

**pre:** R->notEmpty()

**pre:** K->notEmpty()

**pre:** A-> notEmpty()

*--initialisation des ensembles abstraits*

**post:** Guest=G

**post:** Room=R

**post:** Key=K

**post:** Administrator=A

*-- aucune réservation enregistrée : owns:=∅*

**post act1:** owns->isEmpty()

*-- cartes non encore distribuées aux clients : cards:=∅*

 **post act2:** cards->isEmpty()

*-- les clés déjà utilisées sont celles initialement affectées aux chambres : issued:=ran(f)*

**post act3:** issued=f->range()

*-- clés courantes pour les chambres = affectation initiale f des clés aux chambres : currk:=f*

**post act4:** currk=f

*-- les clés mémorisées par les serrures = f : roomk:=f*

**post act5:** roomk=f

*-- cartes non encore distribuées aux personnels : cards_adm:=∅*

**post act6:** cards_adm->isEmpty()

*-- Gestion des réservations*

**context** Hotel:: check_in3(g: GUEST, r: ROOM, c: Pair(Key, Key), a: ADMINISTRATOR)

*-- r est une chambre effective : r∈ROOM*

**pre grd2:** Room->includes(r)

*-- r non encore réservée : r∉dom(owns)*

**pre grd3:** (owns->domain())->excludes(r)

*-- c est une carte effective : c∈CARD*

**pre grd4:** Card->includes(c)

*-- carte appropriée à la chambre r : first(c)=currk(r)*

**pre grd5:** (first->imageElt(c))=(currk->imageElt(r))

*-- la clé seconde ne doit pas être utilisée; éviter l'accès multiple à la même chambre : second(c) ∉ issued*

**pre grd6:** issued->excludes(second->imageElt(c))

*-- la clé seconde ne doit pas être une clé courante pour d'autres chambres ; éviter l'accès du client à multiple chambres : second(c)∉ ran(currk)*

**pre grd7:** (currk->range())->excludes(second->imageElt(c))

*-- c non associée à aucun client : c∉dom(cards)*

**pre grd8:** (cards->domain())->excludes(c)

*--la clé enregistrée par la serrure de la chambre doit être la clé courante associée à cette chambre : roomk(r)=currk(r)*

**pre grd9:** roomk->imageElt(r)=(currk->imageElt(r))

*-- a est un administrateur effectif : a∈ADMINISTRATOR*

**pre grd10:** Administrator->includes(a)

*-- a est l'administrateur de la chambre r : owns_adm(r)=a*

**pre grd11:** owns_adm->imageElt(r)=a

-- *c non associée à aucun personnel : c∉dom(cards_adm)*

**pre grd12:** (cards_adm->domain())->excludes(c)

-- *g devient le propriétaire de r : owns(r)=g*

**post act1:** owns->imageElt(r)=g

-- *la clé seconde de c est marquée utilisée : issued:=issued∪{second(c)}*
**post act2:** issued=issued@pre->including(second->imageElt(c))

-- *la carte c est servie au client g : cards(c):=g*
**post act3:** cards->imageElt(c)=g

-- *la clé seconde de c est enregistrée comme clé courante de r : currk(r):=second(c)*

**post act4:** currk->imageElt(r)=second->imageElt(c)

-- *c est aussi servie à l'administrateur a de r : cards_adm(c):=a*
**post act5:** cards_adm->imageElt(c)=a

-- *Terminaison des réservations*

**context** Hotel:: check_out3(g: GUEST, r: ROOM, c: Pair(Key, Key))

-- *contraintes implicites*

**pre:** Guest->includes(g)

**pre:** Room->includes(r)

**pre:** Card->includes(c)

-- *r doit être réservée par g : r↦g∈owns*
**pre grd1:** owns->imageElt(r)=g

-- *g doit être le propriétaire de c : c↦g∈cards*
**pre grd2:** (cards->imageElt(c))->includes(g)

-- *c est à la propriété d'un administrateur : c∈dom(cards_adm)*

**pre grd3:** (cards_adm->domain())->includes(c)

-- *réservation annulée:* <u>owns:=owns\\{r↦g}</u>

**post act1: let** cpl0:Pair(Room, Guest)=Pair[] **in**

    owns=owns@pre->excluding(cpl0.make(r,g))

-- *carte retirée du client :* <u>cards:=cards\\{c↦g}</u>

**post act2: let** cpl1:Pair(Guest, Pair(Key,Key))=Pair[] **in**

    cards=cards@pre->excluding(cpl1.make(c,g))

-- *carte retirée de l'administrateur approprié :* <u>cards adm:={c}◁cards adm</u>

**post act3:** cards_adm=cards_adm@pre->soustractionDomain(c)

-- *Client entrant dans sa chambre avec changement de clé (pour la première fois)*

**context** Hotel:: enter_room_change3(r: ROOM, c: Pair(Key,Key))

-- *c est retenue par un client :* <u>c∈dom(cards)</u>

**pre grd1:** (cards->domain())->includes(c)

-- *premier accès à la chambre r; la clé enregistrée par la serrure de r doit être celle du client précédent :* <u>roomk(r)=first(c)</u>

**pre grd2:** roomk->imageElt(r)=first->imageElt(c)

-- *la clé seconde ne doit pas être enregistrée dans une autre serrure :* <u>second(c)∉ran(roomk)</u>

**pre grd3:** (roomk->range())->excludes(second->imageElt(c))

-- *chambre réservée :* <u>r∈dom(owns)</u>

**pre grd4:** (owns->domain())->includes(r)

-- *changer la clé enregistrée par la serrure en clé seconde de c :* <u>roomk(r):=second(c)</u>

**post act1:** (roomk->imageElt(r))=second->imageElt(c)

-- *Client entrant dans sa chambre sans changement de clé*

**context** Hotel:: enter_room_normal3(r: ROOM, c: Pair(Key, Key))

-- *c est retenue par un client : c∈dom(cards)*

**pre grd1:** (cards->domain())->includes(c)

-- *chambre réservée : r∈dom(owns)*

**pre grd2:** (owns->domain())->includes(r)

-- *le second accès à r ou plus ; clé de la serrure =clé du client actuel : roomk(r)=second(c)*

**pre grd3:** (roomk->imageElt(r))=second->imageElt(c)

-- *état après= état avant*

**post skip:** owns=owns@pre **and** cards=cards@pre **and** issued=issued@pre **and** currk=currk@pre **and** roomk=roomk@pre **and** cards_adm=cards_adm@pre

-- *Client quittant sa chambre*

**context** Hotel:: leave_room3(r: ROOM, c: Pair(Key, Key))

-- *chambre réservée : r∈dom(owns)*

**pre grd1:** (owns->domain())->includes(r)

-- *c est retenue par un client : c∈dom(cards)*

**pre grd2:** (cards->domain())->includes(c)

-- *le client a déjà entré à la chambre : roomk(r)=second(c)*

**pre grd3:** (roomk->imageElt(r))=second->imageElt(c)

-- *état après= état avant*

**post skip:** owns=owns@pre **and** cards=cards@pre **and** issued=issued@pre **and** currk=currk@pre **and** roomk=roomk@pre **and** cards_adm=cards_adm@pre

*-- Personnel entrant dans une chambre sans changement de clé*

**context** Hotel:: enter_room_normal_adm(r: ROOM, c: Pair(Key, Key))

*-- c est retenue par un personnel : c∈dom(cards_adm)*

**pre grd1:** (cards_adm->domain())->includes(c)

*-- chambre réservée : r∈dom(owns)*

**pre grd2:** (owns->domain())->includes(r)

*-- le second accès à r ou plus ; clé de la serrure =clé du client actuel : roomk(r)=second(c)*

**pre grd3:** (roomk->imageElt(r))=second->imageElt(c)

*-- état après= état avant*

**post skip:** owns=owns@pre **and** cards=cards@pre **and** issued=issued@pre **and** currk=currk@pre **and** roomk=roomk@pre **and** cards_adm=cards_adm@pre

*-- Personnel quittant sa chambre*

**context** Hotel:: leave_room_adm(r: ROOM, c: Pair(Key, Key))

*-- chambre réservée : r∈dom(owns)*

**pre grd1:** (owns->domain())->includes(r)

*-- c est retenue par un personnel : c∈dom(cards_adm)*

**pre grd2:** (cards_adm->domain())->includes(c)

*-- le client a déjà entré à la chambre : roomk(r)=second(c)*

**pre grd3:** (roomk->imageElt(r))=second->imageElt(c)

*-- état après= état avant*

**post skip:** owns=owns@pre **and** cards=cards@pre **and** issued=issued@pre **and** currk=currk@pre **and** roomk=roomk@pre **and** cards_adm=cards_adm@pre

## 4.5 Règles de raffinement d'UML/EM-OCL vers UML/OCL

En partant d'un diagramme de classes décrit en UML/EM-OCL et en appliquant un ensemble de règles de raffinement -proposées dans cette section-, nous pouvons aboutir à un diagramme de classes en UML/OCL décrivant les concepts métier de l'application à traiter. Ces règles de raffinement sont :

(1) Un *attribut de type Set(DataType)* peut être raffiné par une **_classe_** au sens OO

- ➢ **Application :** Guest, Room, Key, Administrator

(2) Un *attribut* qui est utilisé comme ensemble de départ ou d'arrivée pour un autre attribut de type fonction ou relation binaire ou comme paramètre pour des méthodes (et qui a une existence physique) est traduit en une **_classe OO_**

- ➢ **Application** : L'attribut Card est raffiné par une classe

(3) Un *attribut* de type *relation binaire* (totale, partielle, injective, surjective, bijective) ou de type *fonction* (totale, partielle, injective, surjective, bijective) et qui n'est pas utilisé comme ensemble de départ ou d'arrivée ou comme paramètre pour des méthodes est traduit en **_association_** au sens OO (voir **Table 4.4**)

- ➢ **Application :** Les *attributs* first, second, f, owns_adm, owns, cards, currk, roomk et cards_adm sont traduits en associations entre les classes correspondantes

(4) Les **_multiplicités_** des associations sont traduites à partir de la nature des *attributs* (voir **Table 4.4**)

(5) Un *attribut de type Set(Data Type)* lié à l'implémentation peut être affiné par une **_association_** au sens OO

- ➢ **Application** : l'attribut issued est raffiné en association entre Hotel et Key

| Typage en UML/EM-OCL | Typage en UML/OCL |
|---|---|
| x : BinaryRelation(A,B) | A *──x──* B |
| x : PartialFunction(A,B) | A *──x──0..1 B |
| x : TotalFunction(A,B) | A *──x──1 B |
| x : PartialInjective(A,B) | A 0..1──x──0..1 B |
| x : TotalInjective(A,B) | A 0..1──x──1 B |
| x : PartialSurjective(A,B) | A 1..*──x──0..1 B |
| x : TotalSurjective(A,B) | A 1..*──x──1 B |
| x : PartialBijective(A,B) | A 1──x──0..1 B |
| x : TotalBijective(A,B) | A 1──x──1 B |

**Table 4.4 :** *Correspondances entre le typage en UML/EM-OCL et le typage en UML/OCL*

## 4.6  Diagramme de classes en UML/OCL de l'application SCEH

Le diagramme de classes de la **figure 4.3** est un modèle raffiné du modèle UML/EM-OCL de notre application SCEH en appliquant les règles de raffinement proposées dans **4.5**.

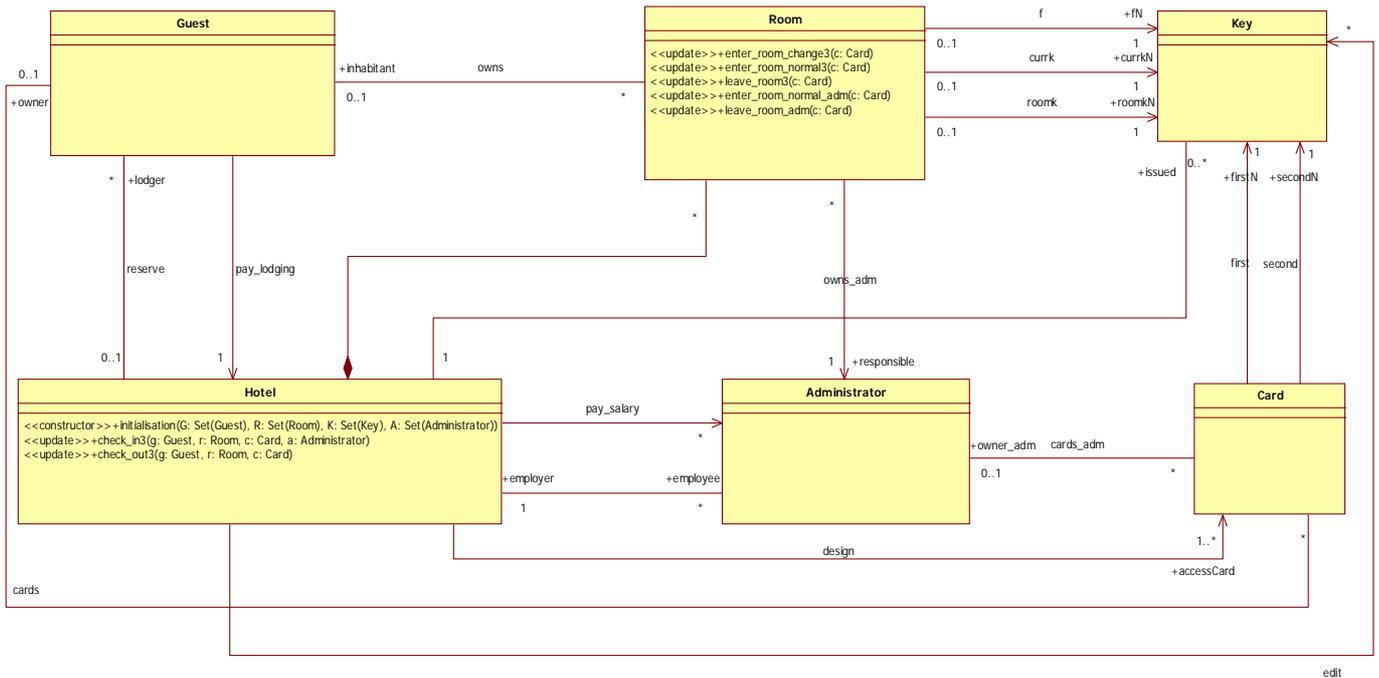

**Figure 4.2 :** *Un modèle raffiné en UML/OCL du modèle SCEH en UML/EM-OCL*

### 4.6.1 Explications

Le diagramme de classe UML/OCL du modèle SCEH est constitué de six classes obtenues et justifiées en raffinant le modèle UML/EM-OCL. En fait, nous avons dispersé les responsabilités de la classe Hotel du modèle UML/EM-OCL sur plusieurs classes UML/OCL. Les classes constitutives du modèle raffiné UML/OCL sont :

- **Hotel** : C'est la classe origine de toutes les autres classes. Elle résulte d'un raffinement de la classe abstraite Hotel par élimination de tous ses attributs et des opérations qui ne la concernent pas. Cette classe est reliée via des associations à toutes les autres classes.

- **Room** : Provenant d'un attribut de type Set(ROOM). Elle n'a pas d'existence propre. Ceci justifie la relation de composition entre Hotel et Room.

- **Guest** : Modélise le concept client *locataire* d'une *chambre* de l'*hôtel*. Les relations conceptuelles fortes entre Guest et respectivement Hotel et Room sont traduites par des associations UML.

- **Administrator** : Modélise le concept personnel.

- **Key**: Modélise la notion de clé électronique. Une telle notion est fondamentale pour la gestion des cartes et pour les chambres, ce qui justifie les associations entre Room et Key.

- **Card** : Le concept carte magnétique est extrait du modèle abstrait et réalisé par une classe UML. Ce choix est justifié en se référent à la *règle (2)*. Une carte est constituée de deux clés nommées first et second : ceci est modélisé par les deux *associations first et second entre les classes Card et key*.

L'élaboration de la partie dynamique de chaque classe est basée sur la distribution des méthodes issues de la classe Hotel du modèle UML/EM-OCL. Une telle répartition est basée principalement sur la *sémantique* de chaque méthode. Par exemple, l'action d'entrer dans une chambre (sans ou avec changement de clé) concerne réellement et essentiellement une instance de type Room. Et par conséquent, les méthodes enter_room_change3, enter_room_normal3 et enter_room_normal_adm doivent appartenir à la classe Room.

### 4.6.2 Contraintes attachées au modèle UML/OCL

Les contraintes OCL liées à la classe Hotel sont :

-- *Propriétés invariantes du système*

**context** Hotel

--propriétés des ensembles en Event-B : un ensemble abstrait est par défaut <u>non vide</u>

**inv not_empty_guest** :  self.lodger->notEmpty()

**inv not_empty_room** :  self.room->notEmpty()

**inv not_empty_key** :  self.key->notEmpty()

**inv not_empty_admin** :  self.employee->notEmpty()

-- l'ensemble des clés utilisées est inclus dans l'ensemble des clés potentielles : <u>issued⊆KEY</u>

**inv inv1_2:** self.key->includesAll(self.issued)

*-- Initialisation du système*

**context** Hotel:: initialisation(G: Set(Guest), R: Set(Room), K: Set(Key), A: Set(Administrator))

*--propriétés venant du niveau supérieur : un ensemble est par défaut <u>non vide</u>*

**pre:** G->notEmpty()

**pre:** R->notEmpty()

**pre:** K->notEmpty()

**pre:** A-> notEmpty()

*--initialisation des ensembles abstraits*

**post:** self.lodger=G

**post:** self.room=R

**post:** self.key=K

**post:** self.employee=A

*-- aucune réservation enregistrée : <u>owns:=∅</u>*

**post act1:** self.room->**forAll** (r: Room|r.inhabitant->isEmpty())

*-- cartes non encore distribuées aux clients : <u>cards:=∅</u>*

 **post act2:** self.accessCard->**forAll** (c:Card|c.owner->isEmpty())

*-- les clés déjà utilisées sont celles initialement affectées aux chambres : <u>issued:=ran(f)</u>*

**post act3:** self.issued=self.room.fN

*-- clés courantes pour les chambres = affectation initiale f des clés aux chambres : <u>currk:=f</u>*

**post act4:** self.room.currkN=self.room.fN

*-- les clés mémorisées par les serrures = f : <u>roomk:=f</u>*

**post act5:** self.room.roomkN=self.room.fN

*-- cartes non encore distribuées aux personnels : <u>cards_adm:=∅</u>*

**post act6:** self.accessCard->**forAll** (c: Card|c.owner_adm->isEmpty())

*-- Gestion des réservations*

**context** Hotel:: check_in3(g: Guest, r: Room, c: Card, a: Administrator)

*-- r est une chambre effective :* <u>r∈ROOM</u>

**pre grd2:** self.room->includes(r)

*-- r non encore réservée :* <u>r↦g∉owns</u>

**pre grd3:** self.lodger->**forAll** (g1: Guest|g1.room->excludes(r))

*-- c est une carte effective :* <u>c∈CARD</u>

**pre grd4:** self.card->includes(c)

*-- carte appropriée à la chambre r :* <u>first(c)=currk(r)</u>

**pre grd5:** c.firstN=r.currkN

*-- la clé seconde ne doit pas être utilisée; éviter l'accès multiple à la même chambre :* <u>second(c) ∉ issued</u>

**pre grd6:** self.issued->excludes(c.secondN)

*-- la clé seconde ne doit pas être une clé courante pour d'autres chambres ; éviter l'accès du client à multiple chambres :* <u>second(c)∉ ran(currk)</u>

**pre grd7:** self.room->**forAll** (r1: Room|r1.currkN<>c.secondN)

*-- c non associée à aucun client :* <u>c∉dom(cards)</u>

**pre grd8:** c.owner->isEmpty()

*--la clé enregistrée par la serrure de la chambre doit être la clé courante associée à cette chambre :*
<u>roomk(r)=currk(r)</u>

**pre grd9:** r.roomkN=r.currkN

*-- a est un administrateur effectif :* <u>a∈ADMINISTRATOR</u>

**pre grd10:** self.employee->includes(a)

*-- a est l'administrateur de la chambre r : <u>owns_adm(r)=a</u>*

**pre grd11:** r.responsible=a

*-- c non associée à aucun personnel : <u>c∉dom(cards_adm)</u>*

**pre grd12:** c.owner_adm->isEmpty()

*-- g devient le propriétaire de r : <u>owns(r)=g</u>*

**post act1:** r.inhabitant=g

*-- la clé seconde de c est marquée utilisée : <u>issued:=issued∪{second(c)}</u>*

**post act2:** self.issued=self.issued@pre->including(c.secondN)

*-- la carte c est servie au client g : <u>cards(c):=g</u>*

**post act3:** c.owner=g

*-- la clé seconde de c est enregistrée comme clé courante de r : <u>currk(r):=second(c)</u>*

**post act4:** r.currkN=c.secondN

*-- c est aussi servie à l'administrateur a de r : <u>cards_adm(c):=a</u>*

**post act5:** c.owner_adm=a

*-- Terminaison des réservations*

**context** Hotel:: check_out3(g: Guest, r: Room, c: Card)

*--propriétés implicites*

**pre:** self.lodger->includes(g)

**pre:** self.room->includes(r)

**pre:** self.card->includes(c)

*-- r doit être réservée par g : <u>r↦g∈owns</u>*

**pre grd1:** r.inhabitant=g

*-- g doit être le propriétaire de c : <u>c↦g∈cards</u>*

**pre grd2:** c.owner=g

-- *c est à la propriété d'un administrateur :* <u>c∈dom(cards_adm)</u>

**pre grd3:** c.owner_adm->notEmpty()

-- *réservation annulée :* <u>owns:=owns\\{r↦g}</u>

**post act1:** r.inhabitant=r.inhabitant@pre->excluding(g)

-- *carte retirée du client :* <u>cards:=cards\\{c↦g}</u>

**post act2:** g.card=g.card@pre->excluding(c)

-- *carte retirée de l'administrateur approprié :* <u>cards_adm:={c}⩤cards_adm</u>

**post act3:** self.employee.card= self.employee.card@pre->excluding(c)

Les contraintes OCL liées à la classe Card sont :

**context** Card

-- *les deux clés d'une carte sont différentes :* <u>∀c·(c∈CARD⇒first(c)≠second(c))</u>

**inv axm1_4:** self.firstN<>self.secondN

Les contraintes OCL liées à la classe Room sont :

-- *Client entrant dans sa chambre avec changement de clé*

**context** Room:: enter_room_change3(c: Card)

-- *c est retenue par un client :* <u>c∈dom(cards)</u>

**pre grd1:** self.inhabitant.card->includes(c)

-- *premier accès à la chambre r; la clé enregistrée par la serrure de r doit être celle du client précédent :* <u>roomk(r)=first(c)</u>

**pre grd2:** self.roomkN=c.firstN

-- *la clé seconde ne doit pas être enregistrée dans une autre serrure :* <u>second(c)∉ran(roomk)</u>

**pre grd3:** self.roomkN->excludes(c.secondN)

-- *chambre réservée : r∈dom(owns)*

**pre grd4:** r.inhabitant ->notEmpty()

-- *changer la clé enregistrée par la serrure en clé seconde de c : roomk(r):=second(c)*

**post act1:** self.roomkN=c.secondN

-- *Client entrant dans sa chambre sans changement de clé*

**context** Room:: enter_room_normal3(c: Card)

-- *c est retenue par un client : c∈dom(cards)*

**pre grd1:** self.inhabitant.card->includes(c)

-- *chambre réservée : r∈dom(owns)*

**pre grd2:** r.inhabitant->notEmpty()

-- *le second accès à r ou plus ; clé de la serrure =clé du client actuel (k2) : roomk(r)=second(c)*

**pre grd3:** self.roomkN=c.secondN

-- *état après= état avant : owns=owns@pre and cards=cards@pre and issued=issued@pre and currk=currk@pre and roomk=roomk@pre and cards_adm=cards_adm@pre*

**post skip:** self. inhabitant=self.inhabitant@pre **and** self.inhabitant.card= self.inhabitant.card@pre **and** self.hotel.issued= self.hotel.issued@pre **and** self.currkN= self.currkN@pre **and** self.roomkN=self.roomkN@pre **and** self.responsible.card= self.responsible.card@pre

-- *Client quittant sa chambre*

**context** Room:: leave_room3(c: Card)

-- *chambre réservée : r∈dom(owns)*

**pre grd1:** r.inhabitant->notEmpty()

-- c est retenue par un client : *c∈dom(cards)*

**pre grd2:** self.inhabitant.cardsN->includes(c)

-- le client a déjà entré à la chambre : *roomk(r)=second(c)*

**pre grd3:** self.roomkN=c.secondN

-- état après= état avant  : *owns=owns@pre and cards=cards@pre and issued=issued@pre and currk=currk@pre and roomk=roomk@pre and cards_adm=cards_adm@pre*

**post skip:** self. inhabitant=self.inhabitant@pre **and** self.inhabitant.card= self.inhabitant.card@pre **and** self.hotel.issued= self.hotel.issued@pre **and** self.currkN= self.currkN@pre **and** self.roomkN=self.roomkN@pre **and** self.responsible.card= self.responsible.card@pre

-- Personnel entrant dans une chambre sans changement de clé

**context** Room:: enter_room_normal_adm(c:Card)

-- c est retenue par un personnel : *c∈dom(cards_adm)*

**pre grd1:** self.responsible.card->includes(c)

-- chambre réservée : *r∈dom(owns)*

**pre grd2:** r.inhabitant->notEmpty()

-- le second accès à r ou plus ; clé de la serrure =clé du client actuel : *roomk(r)=second(c)*

**pre grd3:** self.roomkN=c.secondN

-- état après= état avant

**post skip:** self. inhabitant=self.inhabitant@pre **and** self.inhabitant.card= self.inhabitant.card@pre **and** self.hotel.issued= self.hotel.issued@pre **and** self.currkN= self.currkN@pre **and** self.roomkN=self.roomkN@pre **and** self.responsible.card= self.responsible.card@pre

*-- Personnel quittant sa chambre*

**context** Room:: leave_room_adm(c: Card)

*-- chambre réservée : <u>r∈dom(owns)</u>*

**pre grd1:** r.inhabitant->notEmpty()

*-- c est retenue par un personnel : <u>c∈dom(cards_adm)</u>*

**pre grd2:** self.responsible.card->includes(c)

*-- le client a déjà entré à la chambre : <u>roomk(r)=second(c)</u>*

**pre grd3:** self.roomkN=c.secondN

*-- état après= état avant*

**post skip:** self. inhabitant=self.inhabitant@pre **and** self.inhabitant.card= self.inhabitant.card@pre **and** self.hotel.issued= self.hotel.issued@pre **and** self.currkN= self.currkN@pre **and** self.roomkN=self.roomkN@pre **and** self.responsible.card= self.responsible.card@pre

## 4.7 Raffinement du modèle SCEH en UML/OCL par des classes intermédiaires

Le modèle précédent de la **figure 4.2** peut être raffiné par un autre modèle UML/OCL en lui ajoutant des *classes intermédiaires* dites encore *helpers* (voir **figure 4.3**). Ceci permet d'obtenir un diagramme final raffiné pouvant être traduit directement en une implémentation en langage Eiffel ou Java. Ainsi, les classes helpers introduites sont :

- La classe **Reservation** : la relation existante entre un hôtel (objet de type Hotel) et ses clients (objets de type Guest) peut être soit de réservation soit de payement de logement. La relation de réservation peut être explicitée via une classe Reservation apportant des informations sur les réservations des chambres dans l'hôtel. Ceci est assimilé à un raffinement de l'association « reserve » entre Guest et Hotel.

- La classe **Person** factorise les aspects communs entre les instances de type Guest et celles de type Administrator. Ces aspects concernent entre-autres les droits d'accès aux chambres.

- La classe **Account** modélise l'aspect payement. En effet, le règlement des chambres par les clients ou le versement des salaires des employés nécessite un compte.

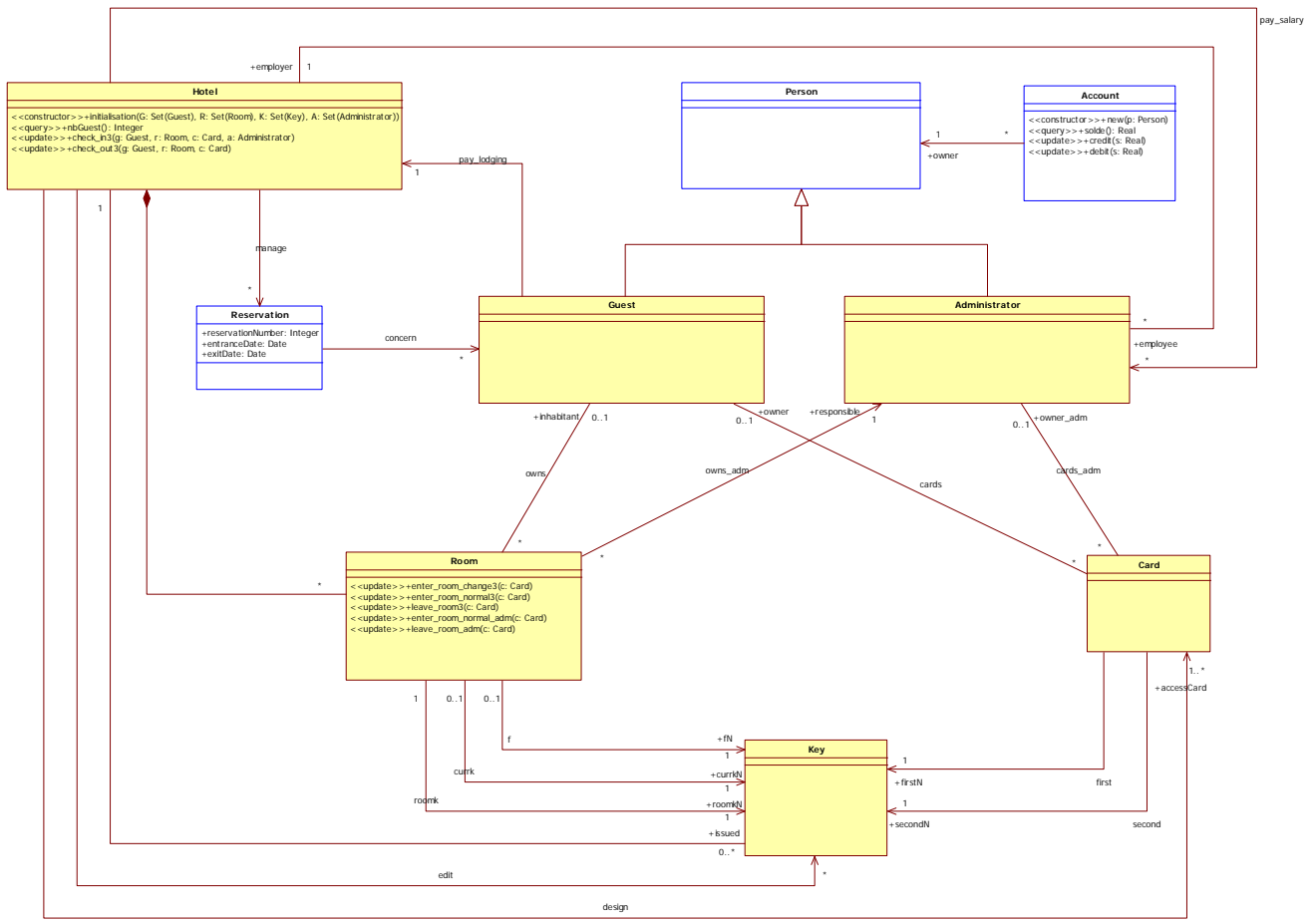

**Figure 4.3 :** *Modèle SCEH raffiné en UML/OCL par des classes helper*

## 4.8 Conclusion

Dans ce chapitre, nous avons présenté d'une façon rigoureuse une transformation de la spécification Event-B du modèle SCEH vers un diagramme de classes UML/OCL en passant par une transformation UML/EM-OCL. A terme, le diagramme UML/OCL obtenu peut être concrétisé en utilisant un langage de programmation supportant la conception par contrats tels que langage Eiffel ou Java couplé à JML.

# *Conclusion*

Dans ce mémoire, nous avons abordé la problématique de développement des logiciels et systèmes complexes sensibles aux erreurs. Pour éviter les erreurs inhérentes au développement de ces logiciels et systèmes, nous avons proposé un processus de développement combinant le formel (Event-B) et le semi-formel (UML/EM-OCL et UML/OCL). La méthode formelle Event-B permet d'obtenir une spécification cohérente et digne de confiance du futur système en utilisant les outils associés à Event-B : générateur d'obligations de preuves, prouveur automatique et interactif, animateur et model-checker. Ensuite, la spécification cohérente et digne de confiance en Event-B du futur système est traduite en UML/OCL en passant par notre extension UML/EM-OCL. Une fois la main est passée à UML/OCL, on suit un processus de développement semi-formel orienté objet en utilisant les techniques associées : patterns de conception de GoF [12], bibliothèques de classes et génération de données de test.

Nous avons appliqué le processus de développement proposé sur une étude de cas Système de Clés Electroniques d'Hôtels (SCEH) décrit dans [3]. Pour y parvenir, dans un premier temps, nous avons structuré le cahier des charges de l'étude de cas SCEH en utilisant les recommandations méthodologiques proposées par J-R Abrial [3]. Dans un deuxième temps, nous avons établi une stratégie de raffinement adaptée à l'étude de cas SCEH. Dans un troisième temps, nous avons spécifié en Event-B le modèle initial et les modèles raffinés de l'étude de cas SCEH. Les propriétés de sûreté et de vivacité –en partie- liées à ces modèles ont été prouvées. Nous avons dû utiliser l'animateur et le model-checker ProB [6] [13] afin de corriger nos modèles Event-B. Dans un quatrième temps, nous avons traduit la spécification abstraite du système SCEH décrite en Event-B vers UML/EM-OCL. Enfin, dans un cinquième temps, nous avons traduit la spécification UML/EM-OCL du système SCEH obtenue vers UML/OCL en utilisant des règles définies par nous-même.

Quant aux perspectives de ce travail, nous pourrions envisager les prolongements suivants :

- Enrichir davantage notre extension EM-OCL afin de couvrir les opérations ensemblistes manquantes telles que identité (id) et produit cartésien ;

- Vérifier davantage les propriétés de vivacité standards ;

- Etablir et vérifier des propriétés de vivacité spécifiques à l'application SCEH en se servant du model-checker ProB ;

- Compléter l'étude de cas SCEH jusqu'à son terme en utilisant les techniques orientées objets associées : patterns de conception, bibliothèques de classes et génération de données de test;

- Automatiser le passage d'Event-B vers UML/EM-OCL ;

- Automatiser le passage d'UML/EM-OCL vers UML/OCL.

Les deux prolongements précédents peuvent être réalisés en utilisant une approche de type IDM [7] [20].

# *Références*